\documentclass[11pt, letterpaper]{article} % for ArXiV
\usepackage[margin=1in]{geometry}
\usepackage{charter}

\usepackage{amsfonts, amsmath, amssymb, amsthm}
\usepackage{bm}
\usepackage{verbatim}
\usepackage{euscript}
\usepackage{graphicx}
\usepackage[usenames,dvipsnames]{xcolor}
\usepackage[normalem]{ulem}
\usepackage{tikz}
\usetikzlibrary{decorations.pathreplacing}
\usetikzlibrary{positioning}
\usetikzlibrary{arrows.meta}
\usetikzlibrary{decorations.pathmorphing}
\usetikzlibrary{calc}

\usepackage{algpseudocode}

\usepackage{xspace}
\usepackage{wrapfig}
\usepackage{url}

\newcommand{\occ}{\mathtt{occ}}
\newcommand{\head}{\mathtt{head}}
\newcommand{\tail}{\mathtt{tail}}
\newcommand{\leaf}{\mathtt{leaf}}
\newcommand{\Dict}{\mathtt{Dict}}
\newcommand{\child}{\mathtt{child}}
\newcommand{\strdepth}{\mathtt{strdepth}}
\newcommand{\lab}{\mathtt{label}}
\newcommand{\slabel}{\mathtt{label}}
\newcommand{\init}{\mathtt{init}}
\newcommand{\weight}{\mathtt{weight}}
\newcommand{\slink}{\mathtt{slink}}
\newcommand{\wlink}{\mathtt{wlink}}
\newcommand{\parent}{\mathtt{parent}}
\newcommand{\lca}{\mathtt{lca}}
\newcommand{\pref}{\mathtt{pref}}
\newcommand{\num}{\mathtt{num}}
\newcommand{\spath}{\operatorname{path}}
\newcommand{\lcp}{\operatorname{LCP}}
\newcommand{\Lowest}{\operatorname{Lowest}}

\newcommand{\cT}{\mathcal{T}}

\newcommand{\cE}{\mathcal{E}}
\newcommand{\cP}{\mathcal{P}}

\newcommand{\gk}[1]{}

\long\def\gktext#1{{\color{magenta}#1}}

\newcommand{\commented}[1]{}
\renewcommand{\arraystretch}{1.3}

\newtheorem{theorem}{Theorem}
\newtheorem{lemma}{Lemma}

\title{Real-Time String Indexing in Optimal Time}
\author{Gregory Kucherov\thanks{LIGM, CNRS, Universit\'e Gustave Eiffel, Marne-la-Vall\'ee, France, \texttt{Gregory.Kucherov@univ-eiffel.fr}} \and
Yakov Nekrich\thanks{Michigan Technological University, USA, \texttt{yakov.nekrich@googlemail.edu}}}

\date{}
\begin{document}
\maketitle
\begin{abstract}
  In this paper we describe the first  fully real-time indexing solution for pattern matching over an arbitrary integer alphabet.  Our data structure spends $O(1)$  worst-case time w.h.p. for processing each new letter appended to $T$ and worst-case $O(|P|+occ)$  time for reporting all occurrences of $P$, where $T$ and $P$ are strings over an  integer alphabet.

  The main idea of our solution is a novel representation of the suffix tree, the path-decomposed suffix tree, which does not store the full topology of the suffix tree but instead decomposes it into paths, each stored in a dictionary that encodes parent--child relationships indirectly.
\end{abstract}

\section{Introduction}
In the string indexing problem, we store a text $T$ in a data structure, so that for an arbitrary 
query string $P$ all occurrences of $P$ in $T$ can be reported efficiently. String indexing is arguably among the most important problems in string algorithms. Numerous algorithmic aspects of string indexes have been studied in the literature, including construction algorithms~\cite{McCreight76,Farach97,FarachFM00,ManberM93,KarkkainenSB06,KimSPP05,KoA05}, compressed indexes~\cite{GrossiV05,FerraginaM05,NavarroM07}, indexes for highly repetitive texts~\cite{GagieNP20}, indexes in external memory~\cite{FerraginaG99}, indexes for approximate pattern matching~\cite{ColeGL04}, document retrieval~\cite{Muthukrishnan02}, and dynamic indexes that support insertions and deletions of symbols~\cite{GuFB94,SahinalpV96,Ferragina97,AlstrupBR00,KempaK22}. 
In the real-time indexing problem, a streaming text must be processed in real time, i.e., with worst-case $O(1)$ time spent on each letter, so that at each moment a pattern-matching query $P$ can be made to the portion of the text received so far. In this paper, we show that the real-time indexing problem can be solved in optimal time. 

\paragraph{Previous Work.}
Maintaining an indexing data structure in \textit{real time} is a long-standing problem that has been studied since the 1970s (see, e.g.,~\cite{Slisenko78}). 
% Whereas a fully real-time construction of suffix trees is not known, several $O(\log\log n)$ solutions have been proposed. 
Early works
on real-time indexing~\cite{Slisenko78,Kosaraju94,AmirN08} are restricted to existence queries over constant-size alphabets.
Amir et al.~\cite{AmirKLL05} presented an online suffix tree construction that processes each letter in $O(\log n)$ worst-case time.  
Breslauer and Italiano \cite{BreslauerI13} presented a \textit{near-real-time} suffix tree construction processing each letter in worst-case $O(\log\log n)$ time for a constant-size alphabet. 
Kopelowitz~\cite{Kopelowitz12} presented a near-real-time suffix tree construction for a general alphabet that processes each character in $O(\log\log n + \log\log \sigma)$ worst-case \textit{expected} time, where $\sigma$ is the alphabet size. This result is achieved by reducing the insertion of each new suffix to predecessor queries on dynamic subsets of an ordered list of suffixes, together with weighted level-ancestor queries.
Fischer and Gawrychowski~\cite{FischerG15} maintain the suffix tree of a growing text in deterministic worst-case $O(\log\log n + \log^2\log\sigma/\log\log\log\sigma)$ time per letter, for $\sigma = n^{O(1)}$, and support pattern matching in $O(|P| + \log^2\log\sigma/\log\log\log\sigma)$ time; their main tool is the wexponential search tree, a weighted variant of exponential search trees~\cite{AnderssonT07}. 
% which are, however, either randomized \cite{6375306} or work for constant-size \cite{BRESLAUER201332} or log-size \cite{kucherov17fullfledged} alphabet. 
A fully real-time indexing solution for string matching over a \textit{constant-size alphabet} has been proposed by the authors in \cite{KucherovN17}. It maintains a complex string index composed of several constituent data structures, each supporting search for patterns of a certain length. However, the requirement of a constant-size alphabet is fundamental for this solution. %Note that all these algorithms maintain the suffix tree for the left-growing text relying on Weiner's right-to-left suffix tree construction \cite{weiner73linear} -- historically the first linear-time suffix tree construction algorithm. 

Despite all these efforts, the real-time indexing problem in its full generality ---~i.e., supporting $O(1)$ time per letter over an arbitrary integer alphabet and $O(|P|+occ)$ time for reporting all $occ$ occurrences of $P$ --- remains an important open problem. The problem appears all the more difficult because all existing approaches to real-time indexing rely on variants of dynamic weighted-level-ancestor or predecessor queries, for which a super-constant lower bound exists~\cite{PatrascuT06}. % OLD citation: \cite{navarro20predecessor}

\paragraph{Our Result.}
In this work, we propose a fully real-time indexing solution for pattern matching over an \textit{arbitrary integer alphabet}. More specifically, our solution spends $O(1)$ worst-case time w.h.p. for processing each new letter appended to $T$ and worst-case $O(|P|+occ)$ time for reporting all occurrences of $P$, where $T$ and $P$ are strings over an arbitrary integer alphabet. Thus, we present the first optimal-time algorithm for real-time string indexing on arbitrary integer alphabets. Our result can be summarized as follows. 

\begin{theorem}\label{thm:main}
Let $T$ be a text over an integer alphabet of arbitrary size that is given
online, one letter at a time, and let $n$ denote the current length of $T$.
On the word RAM, there is a data structure that uses $O(n)$ words of space
and, for any constant $c>0$, processes each new letter in $O(1)$ worst-case
time with probability at least $1-n^{-c}$. At any moment, given a pattern
$P$, it can report all $\mathit{occ}$ occurrences of $P$ in the current text
$T$ in $O(|P|+\mathit{occ})$ worst-case time.
\end{theorem}

Randomization is used only in the dynamic dictionaries, which support lookups in $O(1)$ worst-case time and updates in $O(1)$ time w.h.p.\ (Section~\ref{sec:path-ancestor}).

Our solution, technically involved, relies on several new ideas. The main one is a novel representation of the suffix tree that does not store its full topology but instead decomposes the tree into paths, each stored in a dictionary encoding parent--child relationships indirectly. This allows us to overcome intrinsic limitations of suffix trees with respect to real-time processing. In Section~\ref{sec:overview2} we provide a high-level overview of our solution, which is presented in detail in the subsequent sections.

\section{Extended Overview}
\label{sec:overview2}
\paragraph{Preliminaries.}
We assume familiarity with suffix trees. \textit{Edge labels} are the strings labeling suffix tree edges. A \textit{location} (or locus) in the suffix tree is either an explicit node or an implicit node located on an edge. Thus, each location corresponds to a distinct substring of the string the suffix tree is constructed for. If $\nu$ is a suffix tree node or location, $\lab(\nu)$ denotes the label of $\nu$, which is the concatenation of edge labels on the path from the root to $\nu$, and $|\lab(\nu)|$, also denoted $\strdepth(\nu)$, is the \textit{string depth} of $\nu$.

As in previous works on (near-)real-time indexing~\cite{BreslauerI13,KucherovN17,KopplK26}, % OLD citation: \cite{BRESLAUER201332,kucherov17fullfledged,KopplKucherovCPM26}
we adopt the ``reversed framework'', where the suffix tree is constructed for a left-growing text, i.e., a text where new symbols are prepended to its beginning. For a right-growing text, the same method can be applied to the reversed text, with pattern-matching queries answered on the reversed query string. 
 Weiner's algorithm~\cite{Weiner73} % OLD citation: \cite{weiner73linear}
is the basic tool for online maintenance of the suffix tree for left-growing strings; it is the starting point for subsequent (near-)real-time algorithms, including the one in this work. A key ingredient of Weiner's algorithm is \textit{Weiner links}, letter-indexed links assigned to suffix tree nodes. In more detail, if $\lab(\nu)=s$ for a node $\nu$ and a substring $s$, then a Weiner link of $\nu$ indexed by a letter $a$ points to the locus of $as$ in the tree, provided that $as$ is a substring of the text. If the locus of $as$ is a node, the Weiner link is the reverse of the \textit{suffix link} pointing from the node labeled with $as$ to the node labeled with $s$.
We denote by $\slink(u)$ the target of the suffix link of a node $u$, i.e., if $\lab(u)=as$ for a letter $a$, then $\slink(u)$ is the node $v$ with $\lab(v)=s$. For two nodes $u$ and $v$, $\lca(u,v)$ denotes their lowest common ancestor. Throughout the paper, $n$ denotes the current length of the text.

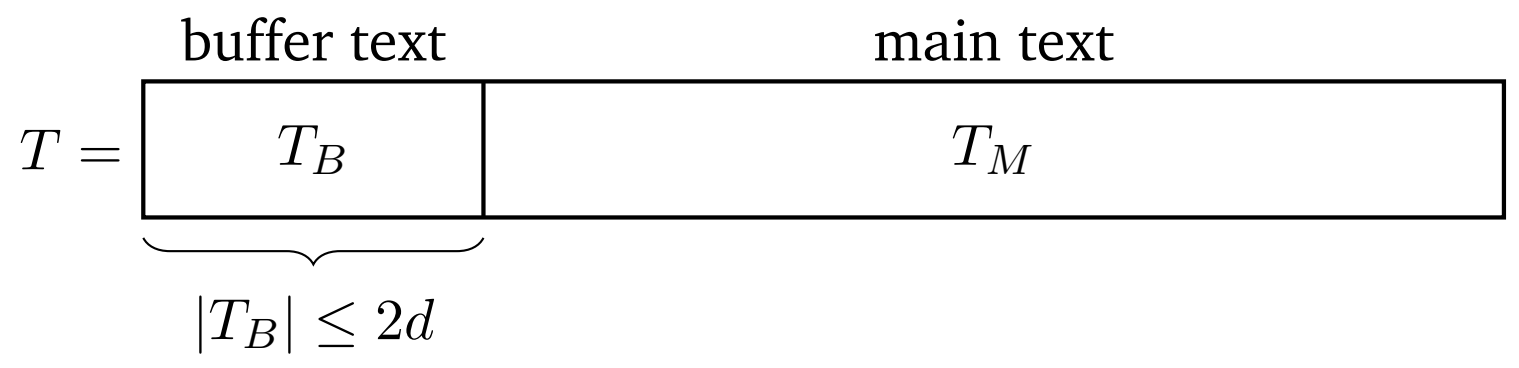
\begin{figure}[t]
\centering
\begin{tikzpicture}[x=0.9cm,y=0.9cm]

% whole text
\draw[thick] (0,0) rectangle (10,1);

% split point
\draw[thick] (2.5,0) -- (2.5,1);

% labels inside boxes
\node at (1.25,0.5) {$T_B$};
\node at (6.25,0.5) {$T_M$};

% text labels above
\node[above] at (1.25,1) {buffer text};
\node[above] at (6.25,1) {main text};

% braces below
\draw[decorate,decoration={brace,mirror,amplitude=5pt}]
  (0,-0.15) -- (2.5,-0.15)
  node[midway,below=7pt] {$|T_B|\le 2d$};

% brace under $S_M$
%\draw[decorate,decoration={brace,mirror,amplitude=5pt}]
%  (2.5,-0.15) -- (10,-0.15)
%  node[midway,below=7pt] {remaining symbols};

% optional label for the whole string
\node[left] at (0,0.5) {$T=$};

\end{tikzpicture}
\caption{\label{fig:buffer-main}Text $T$ is divided into a buffer text $T_B$ and a main text $T_M$.}
\end{figure}

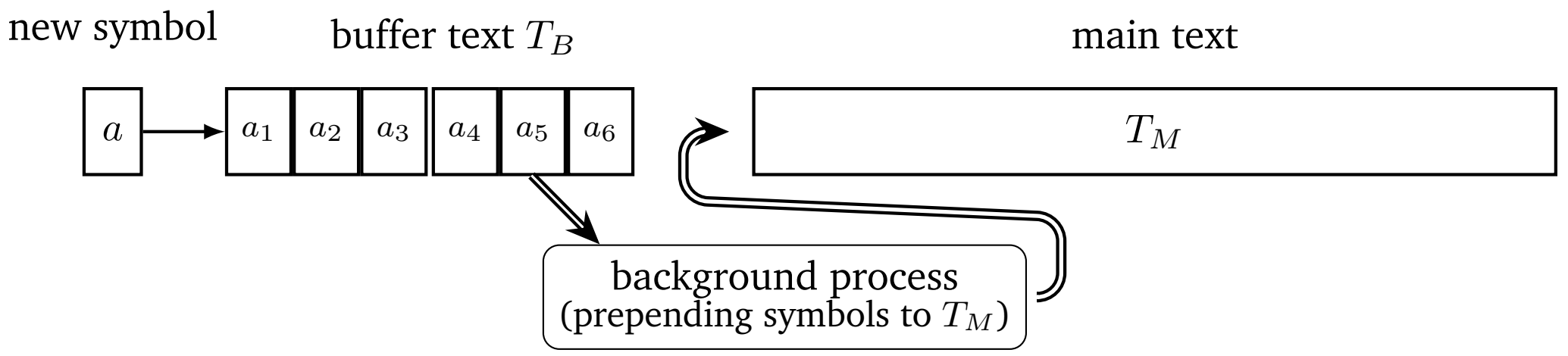
\begin{figure}[t]
\centering
\begin{tikzpicture}[x=0.85cm,y=0.85cm,>=latex,
                    every node/.style={font=\small}]

\tikzset{
  cell/.style={
    draw, thick,
    minimum width=0.5cm,
    minimum height=0.75cm
  }
}

% Incoming symbol
\node[cell] (new) at (-1.5,0) {$a$};
\node[above=2mm of new] {new symbol};

% Buffer
\node[cell, font=\scriptsize] (a1) at (0,0) {$a_1$};
\node[cell, font=\scriptsize,right=0pt of a1] (a2) {$a_2$};
\node[cell, font=\scriptsize,right=0pt of a2] (a3) {$a_3$};
\node[cell, font=\scriptsize,right=1pt of a3] (a4) {$a_4$};
\node[cell, font=\scriptsize,right=0pt of a4] (a5) {$a_5$};
\node[cell, font=\scriptsize,right=0pt of a5] (a6) {$a_6$};

\node[above=2mm] at (2.0,0.4) {buffer text $T_B$};

% Main text -- much longer than the buffer
\node[draw, thick,
      minimum width=7.0cm,
      minimum height=0.75cm] (main) at (9.2,0) {$T_M$};
\node[above=2mm of main] {main text};

% Incoming arrow
\draw[->,thick] (new.east) -- (a1.west);

% Background process
\node[draw,rounded corners,align=center]
      (back) at (5.4,-1.7)
      {background process\\[-1mm]
       \footnotesize (prepending symbols to $T_M$)};

% Part of buffer currently being processed
\draw[double,->,thick,-{Stealth[length=3mm,width=2.5mm]}] (2.8,-0.45) -- ([xshift=5.0mm] back.north west);

% Background work affects main text
%\draw[->,thick] (back.north east) -- (main.south);
%\draw[->,thick]
%  (back.north east)
%  .. controls +(1.0,0.8) and +(-1.0,-0.3) ..
%  ([yshift=-2pt]main.west);
%\draw[->,thick]
%  (back.north east)
%  .. controls +(1.0,0.6) and +(-0.8,-0.8) ..
%  ([xshift=5mm]main.south west);
% \draw[->,thick]
%   ([yshift=-5mm]back.north east)
%   .. controls +(2.5,0.5)  and +(0,-1.0) ..
%   ([xshift=2mm]main.south west);

% \draw[->,thick]
%   ([xshift=3mm]back.east)
%   .. controls +(0.8,0) and +(-0.7,-0.8) ..
%   ([xshift=5mm]main.south west);

% \draw[->,thick]
%   ([xshift=2mm]back.east)
%   .. controls +(0.8,0.0) and +(-0.3,-0.4) ..
%   ([xshift=12mm,yshift=8mm]back.east)
%   .. controls +(0.8,0.5) and +(0,-1.0) ..
%   ([xshift=5mm]main.south west);  

%\draw[->,thick,rounded corners=6pt]
%  ([xshift=2mm]back.east)
%  -- ++(2mm,0)
%  -- ++(0,7mm)
%  -| ([xshift=5mm]main.south west);

\draw[->,double,double distance=1pt, thick,-{Stealth[length=3mm,width=2.5mm]}, rounded corners=6pt]
  ([xshift=1mm]back.east)
  -- ++(2mm,0)
  -- ++(0,7mm)
  -- ([xshift=-6mm,yshift=-6mm]main.west)
  -- ([xshift=-6mm]main.west)
  -- ([xshift=-2mm]main.west);  
  
% explanatory label
%\node[align=center] at (1.1,-1.7)
%      {each new symbol\\triggers some work};

%\draw[->,thick] (0.7,-1.35) -- (0.25,-0.45);

\end{tikzpicture}

\caption{\label{fig:block-move}A new symbol is prepended to the buffer text $T_B$.
At every symbol insertion, some work is performed by the background
process working towards prepending a block of symbols from $T_B$ to the main text $T_M$.
}
\end{figure}

\paragraph{Buffer Text and Main Text.}
The input text $T$ is divided into a short buffer text $T_B$ of length at most $\log n$ and the main text $T_M$. When a new symbol is prepended to $T$, we
first add it to the buffer text. A background process guarantees that the length
of the buffer never exceeds $2d$. The main operation performed by this process
transfers the $d=(1/2)\log n$ rightmost symbols from $T_B$ to $T_M$ (see Figs.~\ref{fig:buffer-main} and~\ref{fig:block-move}). Occurrences in $T_B$ and occurrences that start in $T_B$ and end in $T_M$ are found with a separate, simpler data structure (Sections~\ref{sec:buffertext} and~\ref{sec:marginal}). Thus, in order to obtain
a real-time indexing structure, it is sufficient to maintain an index for the
main text that supports pattern matching in $O(|P|+\occ)$ time and can prepend
a block of $\Theta(d)$ symbols in $O(d)$ total time. Constructing such an index is the main technical contribution of this paper; it relies on several new ideas.

\paragraph{Path-Decomposed Suffix Tree.}
Our data structure for the main text is based on a new representation of the
suffix tree, which we call the \textit{path-decomposed suffix tree} (PST). The set of
nodes of the PST is the same as in the standard suffix tree, but we do not
maintain the complete topology of the tree. In particular, for some nodes, their exact position among the other nodes of the suffix tree is not explicitly
known. This relaxation is crucial: maintaining the complete topology requires dynamic weighted-level-ancestor or predecessor queries, on which previous approaches to real-time indexing typically rely, and for an unbounded alphabet such operations incur a super-constant cost. The PST representation is designed to avoid these
predecessor-type operations while still providing enough information to
navigate the suffix tree and answer pattern-matching queries in optimal time.

In our representation, we decompose the suffix tree into a \textit{big subtree}, consisting of \textit{big nodes}, and
a collection of \textit{small subtrees}. Informally, a node is big if its subtree
contains sufficiently many leaves, while every small subtree (i.e., every connected component consisting
of small nodes only) contains only polylogarithmically many leaves. The complete
topology of each small subtree is maintained explicitly. The big subtree, on the other hand, is represented using its \textit{centroid path decomposition}. Thus,
instead of storing all parent--child relations among big nodes, we organize the
big part of the suffix tree into centroid paths and maintain only the
information needed to (1) move down along a centroid path and (2) navigate between different paths.

The PST can be used to answer pattern-matching queries on the main text. Using the centroid path decomposition, we visit centroid paths starting at the root of the centroid-path tree. We move along each centroid path 
as long as the symbols on its edge labels match the symbols of the pattern. When a mismatch occurs, we try to identify the centroid path on which the search continues. % A complete description is given in Section~\ref{sec:queries-main}. 

\paragraph{Prepending Suffixes and w-Links.}
A major technical challenge is the implementation of the background process that
transfers $d$ symbols from the buffer text to the main text.
When a block of symbols $a_1$,$a_2$,$\ldots$, $a_d$ is moved from $T_B$ to $T_M$, we must update the index for the main text and insert the new suffixes. 
%updating the index when a block of symbols is
%moved from the buffer to the main text.
Suppose that the current main text is $S_{\mathrm{init}}$ and that the symbols
$a_d,a_{d-1},\ldots,a_1$ are prepended to it. We then have to insert the new
suffixes
\[
    a_dS_{\mathrm{init}},\quad
    a_{d-1}a_dS_{\mathrm{init}},\quad
    \ldots,\quad
    a_1a_2\cdots a_dS_{\mathrm{init}}.
\]
To support these insertions, we introduce the concept of \emph{w-links}, which generalize Weiner links to multiple symbols and adapt them to the PST, whose exact topology is not always known.
Note that maintaining exact Weiner links in $O(1)$ worst-case time is not feasible, because it relies on answering predecessor queries. 

% and computation of Weiner links would require to know the tree topology.
Our w-links contain coarser information: while a Weiner link connects two nodes, a w-link connects two centroid paths. A w-link labeled with a string $s$ from a centroid path $h$ to a centroid path $h'$ indicates that $h'$ is the lowest centroid path that contains a location labeled with $s$ followed by the string label of some node on $h$. Importantly, we
store only the source and the destination centroid paths, and not the exact source
node of the corresponding Weiner link.

The new suffixes are inserted one after another. We start at the centroid path
containing the lowest big ancestor of the leaf that holds
$S_{\mathrm{init}}$ and move upward in the centroid-path tree. For each new
suffix, we find the lowest ancestor centroid path that has a w-link labeled with the currently prepended prefix. %As the prefix grows, the source paths of these
Since we only move upward and the centroid-path tree has height $O(\log n)$, we visit $O(\log n)$ centroid paths in total.  
Once such a w-link is found, we use its destination path together with the leaf
representing the previously inserted suffix to determine the exact insertion
position using suffix links and lowest-common-ancestor queries. The new suffix is
then inserted directly (possibly as a child of a new internal node) or passed to a small
subtree. The total time needed to insert $d$ suffixes is thus $O(\log n + d)=O(d)$.

\paragraph{Further Components.}
If a new suffix must be inserted into a small subtree $T(v)$, let $v'=\slink(v)$ and let $R(v')$ be the set of leaves reached by suffix links from the leaves of $T(v)$. We determine the position of the new suffix by predecessor and successor searches for the leaf $l$ that holds the previously inserted suffix in $R(v')$. To perform all these searches in $O(d)$ total time for a block of $d$ suffixes, we have to overcome several challenges; we explain how this is done in Section~\ref{subsec:small-subtr}.

A special case arises when the search reaches the root centroid path and there is no w-link labeled with the current prefix $a_{d-i+1}\cdots a_d$. Then this prefix does not label any location in the big subtree, and hence it occurs fewer than $2\tau$ times in the main text. In this case we look for Weiner links in small subtrees and on leaf centroid paths. If neither search succeeds, then the longest match of each remaining suffix with an existing suffix has length at most $d$, and we find the insertion positions of all remaining suffixes in $O(d)$ total time (Section~\ref{subsec:main-shallow}).

Finally, the PST must be updated as the text grows. When a small subtree becomes too heavy, some of its nodes become big; when the weight of a node on a centroid path leaves the range allowed for its level, the centroid paths below it are rebuilt. The rebuilding is performed gradually, so that only $O(1)$ work is done per insertion.

Throughout the paper, we use the following parameters: $d=\frac12\log n$ is the length of a transferred block, $\tau=\log^5 n$ is the threshold separating small and big nodes, and $\rho=\log^2 n$ determines which symbols are significant in Section~\ref{subsec:small-subtr}.

% I need to re-write it or remove it
% The two main components of the PST therefore use the same principle. For
% pattern matching, we do not store the complete topology of the big subtree;
% instead, centroid paths and local branching information allow us to reconstruct
% the route followed by the query. For updates, we do not store the exact
% starting node of every Weiner link; instead, path-level $w$-links identify the
% relevant region of the tree, and suffix links together with lowest-common-
% ancestor information recover the exact insertion point when needed. This
% allows us to avoid the predecessor operations that arise in standard dynamic
% suffix-tree representations while retaining enough information to support
% optimal pattern matching and efficient batched updates.

\section{Data Structure for Main Text}
\label{sec:maintext}
We start with an overview of our main data structure, the path-decomposed suffix tree (PST). The set of nodes of the PST is the same as in the standard suffix tree. The main (and crucial) difference is that for some nodes in the PST, their exact position in the suffix tree is not known.

We classify PST nodes into big and small. Let $\tau=\log^5 n$. A PST node with at least $2\tau$ leaf descendants is always \textit{big}, and a node with at most $\tau$ leaf descendants is always \textit{small}; nodes with more than $\tau$ and fewer than $2\tau$ leaf descendants can be either small or big. Our algorithm also guarantees that all ancestors of a big node are big nodes. Thus, the PST is partitioned into a \emph{big subtree} and a number of \emph{small subtrees}. The big subtree consists of all big nodes; every small subtree has fewer than $2\tau$ leaves. If a node $u$ is the root node of a small subtree, then its parent is a big node. We can maintain the decomposition of a tree into a big subtree and small subtrees \cite{AlstrupHR98}.

\begin{figure}[t]
\centering
\begin{tikzpicture}[x=0.9cm,y=0.9cm,
  big/.style={circle,fill=black,inner sep=1.4pt},
  pth/.style={line width=2.2pt},
  tree/.style={thin,gray!70!black},
  small/.style={draw=gray!70!black,fill=gray!20,thin},
  lab/.style={font=\footnotesize}]
\def\smalltri#1#2{\fill[small] (#1,#2) -- ++(-0.45,-1.0) -- ++(0.9,0) -- cycle;}
% ---- small subtrees (drawn first)
\draw[tree] (0,0) -- (-1.4,-1.1); \smalltri{-1.4}{-1.1}
\draw[tree] (0,-3.6) -- (-0.6,-4.6); \smalltri{-0.6}{-4.6}
\draw[tree] (0,-3.6) -- (0.6,-4.6); \smalltri{0.6}{-4.6}
\draw[tree] (2.6,-4.6) -- (2.0,-5.6); \smalltri{2.0}{-5.6}
\draw[tree] (2.6,-4.6) -- (3.2,-5.6); \smalltri{3.2}{-5.6}
\draw[tree] (-2.4,-4.6) -- (-2.4,-5.6);\smalltri{-2.4}{-5.6}
\draw[tree] (4.6,-5.6) -- (4.6,-6.6); \smalltri{4.6}{-6.6}
% ---- centroid paths; the edge from parent(head(h)) to head(h) belongs to h
\draw[pth,blue!70!black] (0,0) -- (0,-3.6);
\draw[pth,orange!90!black] (0,-1.2) -- (2.6,-2.2) -- (2.6,-4.6);
\draw[pth,violet] (0,-2.4) -- (-2.4,-3.4) -- (-2.4,-4.6);
\draw[pth,cyan!60!black] (2.6,-3.4) -- (4.6,-4.4) -- (4.6,-5.6);
% ---- big nodes
\foreach \p in {(0,0),(0,-1.2),(0,-2.4),(0,-3.6),(2.6,-2.2),(2.6,-3.4),(2.6,-4.6),
                (-2.4,-3.4),(-2.4,-4.6),(4.6,-4.4),(4.6,-5.6)}
  \node[big] at \p {};
% ---- labels
\node[lab,blue!70!black,left] at (0,-0.6) {$h_1$};
\node[lab,orange!90!black,right] at (2.6,-2.8) {$h_2$};
\node[lab,violet,left] at (-2.4,-4.0) {$h_3$};
\node[lab,cyan!60!black,right] at (4.6,-5.0) {$h_4$};
\node[lab,above] at (0,0) {root};
\node[lab,right] at (2.65,-2.05) {$\head(h_2)$};
\node[lab,left] at (2.55,-4.6) {$\tail(h_2)$};
\fill (3.2,-6.45) circle (1.4pt);
\node[lab,below] at (3.2,-6.6) {$\leaf(h_2)$};
\node[lab,gray!40!black] at (0,-6.1) {small subtrees};
\draw[gray!60!black,-{Latex[length=1.5mm]}] (-0.4,-5.95) -- (-0.55,-5.65);
\draw[gray!60!black,-{Latex[length=1.5mm]}] (0.4,-5.95) -- (0.55,-5.65);
% ---- centroid-path tree
\begin{scope}[xshift=7.6cm,yshift=-1.3cm]
  \node[lab] at (0,0.9) {centroid-path tree};
  \node[circle,draw,inner sep=1pt,lab,blue!70!black] (H1) at (0,0) {$h_1$};
  \node[circle,draw,inner sep=1pt,lab,violet] (H3) at (-0.9,-1.3) {$h_3$};
  \node[circle,draw,inner sep=1pt,lab,orange!90!black] (H2) at (0.9,-1.3) {$h_2$};
  \node[circle,draw,inner sep=1pt,lab,cyan!60!black] (H4) at (0.9,-2.6) {$h_4$};
  \draw (H1) -- (H3); \draw (H1) -- (H2); \draw (H2) -- (H4);
\end{scope}
\end{tikzpicture}
\caption{\label{fig:pst}The PST. Black dots are big nodes, and grey triangles are small subtrees, whose topology is stored explicitly. The big subtree is decomposed into centroid paths $h_1,\ldots,h_4$ (thick edges); every edge of the big subtree belongs to some centroid path, and the edge from $\parent(\head(h))$ to $\head(h)$ belongs to $h$. Each centroid path is drawn in its own color, and the same color is used for the corresponding node of the centroid-path tree on the right. The children of a node $u$ that are not on the centroid path of $u$ are stored in $\child_u$, and for the head of every path we also store its parent. For every centroid path $h$, we store $\head(h)$, $\tail(h)$, $\leaf(h)$, and the dictionary $\Dict_h$ that maps string depths to the nodes of $h$; the order of the nodes on $h$ is not stored explicitly.}
\end{figure}
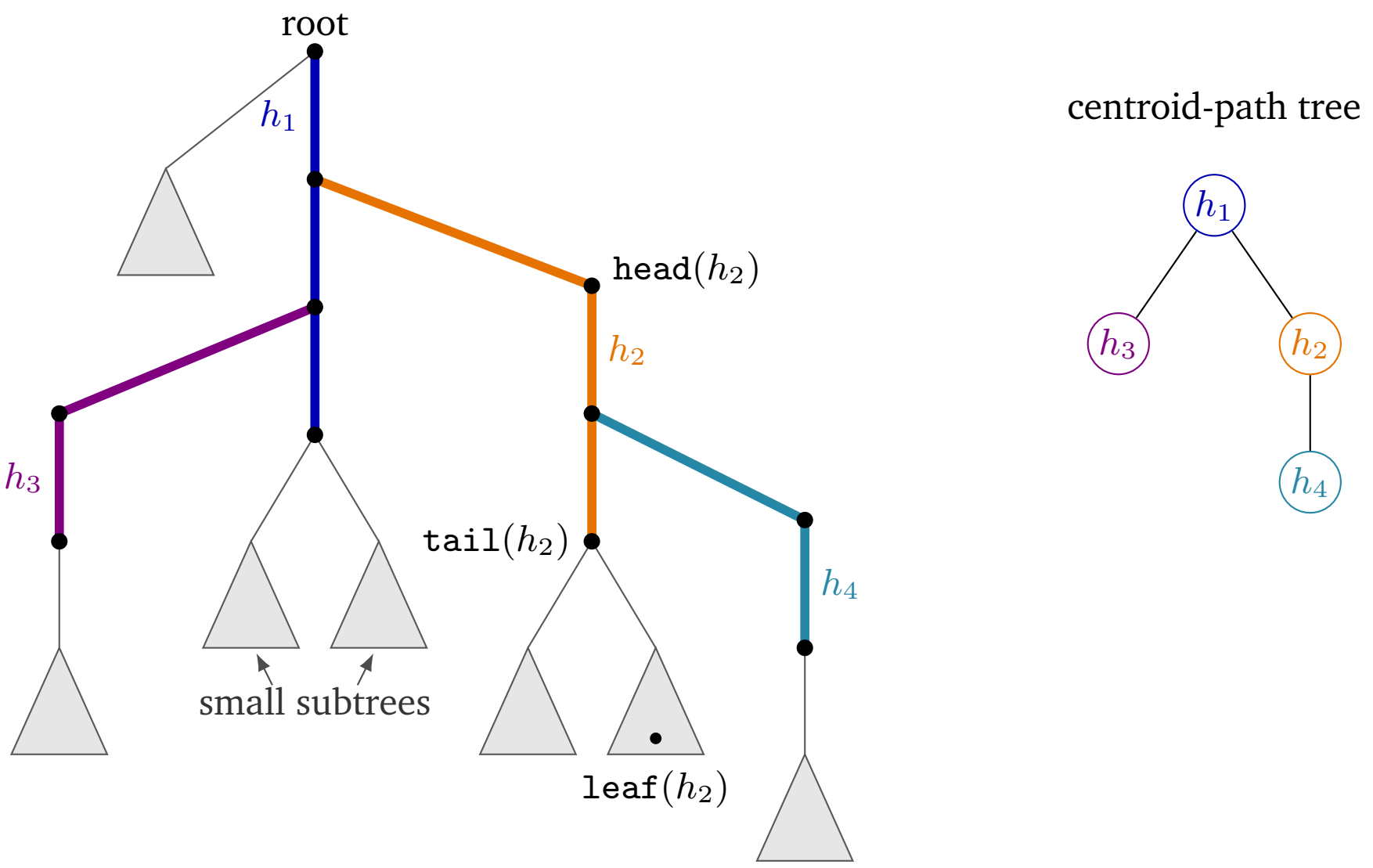

Furthermore, we apply centroid path decomposition~\cite{ColeH05} to the big subtree. Let the \textit{weight} of a node $u$ denote the number of leaves in the subtree rooted at $u$. A level-$i$ centroid path, for $i\ge0$, is a path between a node $u$ and its ancestor $v$ such that:
\begin{enumerate}
\renewcommand{\labelenumi}{(\roman{enumi})}
\item $\weight(u)\ge 2^i\tau$ and $\weight(v)\le (2^{i+1}+2^{i-2})\tau$;
\item the parent of $v$ is on a level-$j$ centroid path for $j\ge i+1$.
\end{enumerate}
\noindent
Every node of the big subtree is assigned to exactly one centroid path; see Fig.~\ref{fig:pst}. 
% NEW FIGURE: illustration of the PST
\gk{So is this some ad-hoc decomposition, not a ``standard'' centroid path decomposition from \cite{doi:10.1137/S0097539796313477}? Yakov: No, not quite}
% $weight(v)\le \frac{n}{2^{i+1}}$ and $weight(u)\ge \frac{n}{2^{i+1}}$.
The centroid path decomposition of the big subtree is stored in the centroid-path tree: the parent of a path $h$ is the path that contains $\parent(\head(h))$. For each centroid path $h$, we maintain its highest and lowest nodes, denoted $\head(h)$ and $\tail(h)$, respectively, as well as an arbitrary leaf descendant of $\tail(h)$, denoted $\leaf(h)$.

We do not always explicitly store parent--child relationships between big nodes of the tree. Navigation in the big subtree is based on centroid paths. For every centroid path $h$, we maintain a dictionary $Dict_h$ such that $Dict_h(x)=u$ iff there is a node $u\in h$ with $\strdepth(u)=x$. Thus, the order of nodes within a centroid path is not explicitly stored.
 We need, however, some additional information in order to navigate between centroid paths. For each node $u\in h$, the dictionary $\child_u$ contains information about all children $u_i$ of $u$ such that $u_i\not\in h$: $\child_u(a)=u_i$ iff the edge from $u$ to $u_i$ is labeled with a string starting with the symbol $a$. We also store the parent $v$ of a node $u$ if $u=\head(h)$ for some centroid path $h$. For each small node $v$ in the PST, we explicitly store all its children as well as its parent $\parent(v)$. In other words, the complete topology of every small subtree is explicitly stored. 
 We remark that the PST is identical to the standard suffix tree, e.g., with respect to the order of leaves. \gk{Is the order of leaves (lexicographic?) relevant? Yakov: I wanted to stress that the tree is the same, we just do not know everything about it.} As explained above, the difference is that the complete topology of the PST is not available to the algorithm. 

\commented{ 
\subsection{To Do: Operations on PST for the Main Text}
1. Maintain centroid paths\\
2. For every node in a small tree, its lowest big ancestor\\
3. Centroid paths: answer wla queries on the centroid path tree\\
4. suffix links (and Weiner links for small tree? Probably just for leaves)\\
5. Maintain the small trees 
}

\section{Main Text: Pattern Matching Queries}
\label{sec:queries-main}
In this section we show how we answer pattern-matching queries using the PST.

\paragraph{Overview.}
To report all occurrences of a query string $P[1..m]$ in $T$, 
we search for $P$ by traversing the centroid-path tree. Starting
from the centroid path containing the root, we compare consecutive symbols of
$P$ with the symbols on the current centroid path. If the pattern continues
along the path, we simply proceed with the comparison. If the pattern leaves
the current path, the dictionary for the path identifies the node $u$ at which the
branching occurs, and the child dictionary of this node determines the next
centroid path from the following pattern symbol. If the search enters a small
subtree, the remaining part of the pattern is searched for using the explicitly
stored topology of that subtree. In this way, the locus of $P$ can be found
in $O(|P|)$ time.

Once the locus of $P$ has been found, all occurrences can be reported by
enumerating the leaves below the locus. The PST does not explicitly store the
complete topology of the big subtree, but the centroid-path representation is
sufficient for this purpose as well. We enumerate the relevant nodes of the
current centroid path and recursively traverse the subtrees that branch away
from it. Every visited node can be charged to a node of the corresponding
ordinary suffix-tree subtree, whose number of internal nodes is linear in its
number of leaves. Consequently, reporting all occurrences takes $O(\occ)$
additional time, and the total query time is $O(|P|+\occ)$.

\paragraph{Finding the Locus of $P$.}
 Let the label $\lab(h)$ of a centroid path $h$ be the concatenation of its edge labels, ordered from top to bottom.
 We remark that $\lab(h)=T[i_h+f_1..i_h+f_2-1]$, where the suffix stored in $\leaf(h)$ is $T[i_h..]$, $f_1=\strdepth(\parent(\head(h)))$ \gk{Shouldn't it be just $f_1=\strdepth(\head(h))$? The convention is that the path edge from head(h) to its parent is included in the path}, and $f_2=\strdepth(\tail(h))$. Hence any symbol on $\lab(h)$ can be accessed in constant time. 
 \begin{lemma}
   \label{lemma:pathlabel}
   For a centroid path $h$ and an integer $k$, we can retrieve the $k$-th symbol on the label of $h$ in time $O(1)$.  
 \end{lemma}
 \begin{proof}
   Let $T[i_h.. n]$ denote the suffix that is assigned to $\leaf(h)$. Let $t$ be the string depth of $\parent(\head(h))$.
   All suffixes in the subtree rooted at $\tail(h)$ are prefixed by string $s\circ\ell$ where $s$ is a string of length $t$ and $\ell$ is the label of $h$. Hence the $k$-th symbol in the label of $h$ is $T[i_h+t+k-1]$.  
 \end{proof}
 We find the locus of $P$ using the centroid path tree: Initially we set $h_c\leftarrow h_R$ where $h_R$ is the root centroid path and $k\leftarrow 1$. Let $T[i_h..]$ be the suffix stored in $\leaf(h_c)$. We compare $P[k]$ with $T[i_h+k-1]$, the symbol at string depth $k$ on the path from the root to $\tail(h_c)$, and increment $k$ until $P[k]\not=T[i_h+k-1]$, or the end of $P$ or of $\lab(h_c)$ is reached. If the end of $P$ is reached (i.e., $k>m$), then we found the locus of $P$; we then identify all nodes on $h_c$ with string depth at least $m$ and their leaf descendants as will be explained below. If $P[k]\not=T[i_h+k-1]$, we find the node $u$ with string depth $k-1$ on $h_c$ using $\Dict_{h_c}$. If the end of $h_c$ is reached, we set $u=\tail(h_c)$. In both cases, we compute $u'=\child_u(P[k])$. If $u$ or $u'$ does not exist, then $P$ does not occur in $T$. If $u'$ is a small node, we search for the locus of $P[k..m]$ in the subtree rooted at $u'$. Otherwise, we set $h_c$ to be the path with $\head(h_c)=u'$ and compare the symbols in $P$ with the symbols on the path from $u$ to $\tail(h_c)$.  
The pseudocode is provided in Fig.~\ref{fig:query1}.

%\begin{lemma}
%  \label{lemma:locus1}
%  Let $u=\head(h)$ for some centroid path $h$. We can find all $N$ leaf descendants of $u$ in $O(N)$ time. 
%\end{lemma}
%\begin{proof}
%  Using $\Dict_h$ we report all nodes $v$ on $h$. \gk{How do we do that? A dictionary does not generally support enumeration.} For every $v$, we report all its children $v'$ such that $v'\not\in h$. For every $v'$ we recursively find all its leaf descendants.
%  To obtain the time bound, we consider the subtree $T(u)$ of the regular suffix tree rooted at $u$.  Every node $v$ or $v'$ visited by our procedure corresponds to an internal node of this subtree. Every  leaf descendant of $u$ corresponds to a leaf in this subtree.  Since the number of leaves and internal nodes in $T(u)$ is bounded by $O(N)$, our procedure runs in $O(N)$ time. 
%\end{proof}

\paragraph{Reporting Occurrences.}
Once the locus $w$ of $P$ in the PST has been identified, we have to report all leaves below $w$. Note that $w$ can be a location on an edge of the suffix tree. If $w$ is in a small subtree, we report the leaves below $w$ by traversing this subtree. Otherwise, $w$ is on the current centroid path $h_c$. We enumerate all nodes on the current centroid path $h_c$. Let $n_h$ denote the total number of nodes on $h_c$. Let $n'_h$ (resp.\ $n''_h$) denote the number of nodes $u$ with $\strdepth(u)< m$ (resp.\ $\strdepth(u)\ge m$). Then $n'_h< m$ and $n''_h=O(\occ)$ where $\occ$ is the number of occurrences of $P$. We then find all nodes $u''\in h_c$ with $\strdepth(u'')\ge m$ in $O(|P|+n''_h)$ time. For every such node $u''$ we examine all children $v$ of $u''$ such that $v\not\in h_c$. % For every $v$ we visit all leaves below $v$.
We then have to visit all leaves below each such $v$. Note that $v$ is either the head of a centroid path or the root of a small subtree. 

Visiting all leaves below the head $v$ of a centroid path $h$ is done by enumerating all nodes of $h$, reporting, for each of them, all its children that are outside $h$ and recursively applying the traversal to them. If $v$ is the root of a small subtree, we directly traverse the tree top-down. 

Every node visited by the procedure is a node of the original suffix tree as well, and their total number is linear in the number of leaves below $v$. Altogether, the procedure takes time $O(|P|+n''_h+\occ)=O(|P|+\occ)$.

\gk{GK: I shortened this part and commented out one Lemma. I think it is simple (and standard) enough and the Lemma is not really needed}

%Let $w_u$ denote the highest node on $h_c$ with $\strdepth(w_u)\ge m$. We consider the conceptual subtree $\cT(w_u)$ of the regular suffix tree rooted at $w_u$. The total number of leaves in this subtree is equal to $\occ$.  Every internal node $u''$ or $v$ visited by our procedure corresponds to an internal node in $\cT(w_u)$.  Since the number of internal nodes in $\cT(w_u)$ does not exceed the number of leaves in $\cT(w_u)$, we visit $O(\occ)$ nodes $u''$ and $v$. Every leaf descendant of $w_u$ is a leaf in $\cT(w_u)$. Every leaf descendant of $w_u$ is a leaf descendant of exactly one node $v$. Hence, by Lemma~\ref{lemma:locus1}, all leaf descendants of all $u''$ can be identified in $O(\occ)$ time.  Hence the total runtime is $O(|P|+n''_h+\occ)=O(|P|+\occ)$. 

 % $h_c\leftarrow h_R$\\
 % $k\leftarrow 1$\\
 % repeat\\
 % $i_h\leftarrow$ starting position of the suffix in $\leaf(h_c)$
 % while  ($P[k]==T[i_h+k]$)  AND ($k\le \min(m, \strdepth(\tail(h_c)))$\\
 % $k\leftarrow k+1$\\

 % if ($P[k]\not=T[i+k]$) OR ($k> \strdepth(\tail(h_c))$)\\
 % $u\leftarrow Dict_{h_c}(i+k-1)$\\
 % $u'\leftarrow \child_u(P[k])$\\
 % if ($u==NULL$) OR ($u'==NULL$)\\
 % Terminate\\
 % else\\
 % $h_c\leftarrow$ central path of $u'$\\

 %until ($k>m$)\\

\newcounter{algline}
\newcommand{\algline}{\thealgline:\quad &}
\newcommand{\nextline}{\stepcounter{algline}\\}

\setcounter{algline}{1}
\begin{figure}[tb]

 \fbox{
	\begin{minipage}{0.9\textwidth}
		\begin{flushleft}
			\renewcommand{\arraystretch}{1.0}
			\begin{tabular}{r@{\quad}l}
				\algline $h_c \leftarrow h_R$ \nextline
				\algline $k \leftarrow 1$ \nextline
				
				\algline \textbf{repeat} \nextline
				\algline \hspace*{1cm}$i_h \leftarrow \text{starting position of the suffix of } \leaf(h_c)$ \nextline
				
				\algline \hspace*{1cm}\textbf{while} 
				$(k \le \min(m,\strdepth(\tail(h_c))))$ \textbf{and} $(P[k]=T[i_h+k-1])$ \nextline
				\algline \hspace*{2cm}$k \leftarrow k+1$ \nextline
				
				\algline \hspace*{1cm}\textbf{if} $k>m$ \nextline
				\algline \hspace*{2cm} \textbf{return} ``match found'' \nextline
				
				\algline \hspace*{1cm}\textbf{else} \nextline
				\algline \hspace*{2cm}$u \leftarrow \Dict_{h_c}(k-1)$ \nextline
				\algline \hspace*{2cm}$u' \leftarrow \child_u(P[k])$ \nextline
				
				\algline \hspace*{2cm}\textbf{if} $(u=\mathtt{NULL})$ \textbf{or} $(u'=\mathtt{NULL})$ \nextline
				\algline \hspace*{3cm}\textbf{return} ``no occurrence'' \nextline
				\algline \hspace*{2cm}\textbf{elseif} ($u'$ is a small node) \nextline
				\algline \hspace*{3cm}search for $P[k..m]$ in the subtree rooted at $u'$; \textbf{return} the result \nextline
				\algline \hspace*{2cm}\textbf{else} \nextline
				\algline \hspace*{3cm}$h_c \leftarrow \text{centroid path of } u'$ \nextline
				\algline \textbf{until} a \textbf{return} statement is executed \nextline
			\end{tabular}
		\end{flushleft}
	\end{minipage}
}
\caption{Query algorithm for finding the locus of $P[1..m]$. \gk{Pseudo-code slightly modified, old version in comments}}
\label{fig:query1}
\end{figure}

\section{Main Text: Prepending Suffixes}
\label{subsec:appendI}
As explained in Section~\ref{sec:overview2}, we run the background process that transfers $d=(1/2)\log n$ symbols $a_d,a_{d-1},\ldots,a_1$ from the buffer text to the main text. 
In this section, we explain how $d$ new suffixes are inserted into the data structure when $d$ new symbols are prepended to the main text $T$. Let $S_{\init}=T[1..n]$ denote the suffix that starts\footnote{To simplify the description, we order symbols in the left-to-right order, so that the leftmost text symbol is the symbol $T[1]$. In a real implementation the symbols can be stored, e.g., in a dynamic array $T_R$ in the right-to-left order so that $T[i]=T_R[n-i+1]$; when a new symbol is prepended to $T$, we add it at the end of $T_R$.}
 at $T[1]$. When new symbols are added, we have to find the positions of suffixes $a_dS_{\init}$, $a_{d-1}a_dS_{\init}$, $\ldots$, $a_1a_2\ldots a_dS_{\init}$ in the PST and insert the corresponding nodes into the PST. Our method is based on the notion of w-links, defined below.

\paragraph{Generalized Weiner Links and w-Links.}
%We will say that there is a w-link labeled with a symbol $a$ from a path $h$ to a path $h'$ if there is at least one Weiner link from $u\in h$ to $v\in h'$ labeled with $a$ and $v$ is a node in a big subtree.
We generalize the notion of Weiner links to strings of length $p\ge 2$.
%We will say that there is a Weiner link from a node $u$ to a node $v$ labeled with a string $s=a_1a_2\ldots a _p$ for some $p\ge 2$ iff there is a Weiner link from $u$ to some $u'$ that is labeled with $a_2\ldots a_p$ and a Weiner link from $u'$ to $v$ that is labeled with $a_1$.
We will say that there is a Weiner link from a node $u$ to a location $\nu$ labeled with a string $s=a_1a_2\ldots a _p$ for some $p\ge 1$ iff $\slabel(\nu)=s\cdot \slabel(u)$. In our data structure, we do not store generalized Weiner links for $p>1$ because maintaining them would be too costly. However, such conceptual Weiner links are used in some proofs.

For a centroid path \(h\) and a nonempty string \(s\), we define
\(\operatorname{wlink}(h,s)\) to be the lowest centroid path \(h'\)
such that there is a Weiner link labeled with \(s\) from some node
\(u\in h\) to a location \(\nu\in h'\).
Here, lowest means deepest in the centroid-path tree.
If no such path exists, we set
\(\operatorname{wlink}(h,s)=\mathrm{NULL}\).
Intuitively, $\wlink(h,s)=h'$ if $h'$ is the lowest centroid path such that
$\slabel(\nu)=s\cdot \slabel(v)$ for some location $\nu\in h'$ and $v\in h$.
For convenience, we will assume that the edge connecting $\head(h)$ with its parent is on the path $h$. 

%there is a w-link from a path $h$ to a path $h'$ that is marked with $s$ if there are many suffixes that are prefixed with $s\circ \head(h')$ where $s'$ is the string label of $\head(h)$.

\paragraph{Overall Method.}
%Our procedure for inserting the new suffixes is divided into two stages.
Let $l_0$ be the leaf that holds $S_{\mathrm{init}}$;
it was the last leaf created in the previous transferring round of the background
process. 
Let $h_0$ be the centroid path that contains the lowest big ancestor of $l_0$.
We find the lowest ancestor path $h_1$ of $h_0$ (possibly $h_0$ itself), such that there is a w-link from $h_1$ to some centroid path $h'_1$ that is labeled with $a_d$. Once $h'_1$ is known, we insert $a_dS_{\init}$ into the PST using the procedure that will be described below. Next, we look for the lowest ancestor path $h_2$ of $h_1$, such that there is a w-link from $h_2$ that is labeled with $a_{d-1}a_d$. When $h_2$ is found, we insert $a_{d-1}a_dS_{\init}$ into the PST. We continue in the same way until all new suffixes are inserted. At iteration $i$, we find the lowest ancestor path $h_i$ of $h_{i-1}$, such that there is a w-link from $h_i$ that is labeled with $a_{d-i+1}\ldots a_d$ and insert $S_i=a_{d-i+1}\ldots a_dS_{\init}$. The complete insertion procedure is shown in Fig.~\ref{fig:append}. The w-links of every centroid path $h$ are stored in the
w-dictionary of $h$ (Section~\ref{sec:path-ancestor}).
Both when $s_i$ is extended by one symbol and when we
move from $h_i$ to $\parent(h_i)$, the entry for the current
string is found in $O(1)$ time; hence each test
$\wlink(h_i,s_i)=\mathtt{NULL}$ in Fig.~\ref{fig:append}
takes $O(1)$ time.

\setcounter{algline}{1}
\begin{figure}[tb]
  \fbox{
    \begin{minipage}{0.8\textwidth}
      \begin{flushleft}
        \renewcommand{\arraystretch}{0.8}
        \begin{tabular}{r@{\quad}l}
          & \\
          \algline $l_0\leftarrow$ the leaf that holds $S_{\init}$ \nextline
          \algline $s_0$$\leftarrow $ empty string \nextline
          \algline $i\leftarrow 0$ \nextline
          & \\
          \algline $h_0\leftarrow$ centroid path that contains the lowest big ancestor of $l_0$ \nextline
          \algline \textbf{repeat} \nextline
          \algline \hspace*{1cm} $s_{i+1}\leftarrow a_{d-i}s_i$ \nextline
          \algline \hspace*{1cm} $i\leftarrow i+1$ \nextline
          \algline \hspace*{1cm} $h_i\leftarrow h_{i-1}$ \nextline
          \algline \hspace*{1cm} \textbf{while} ($\wlink(h_i,s_i)=\mathtt{NULL}$) \textbf{and} ($h_i\not= h_R$) \nextline
          \algline \hspace*{2cm} $h_i\leftarrow \parent(h_i)$ \nextline
          % \algline  \hspace*{1cm} \textbf{endwhile} \nextline
          \algline \hspace*{1cm} \textbf{if} ($h_i= h_R$) \textbf{and} ($\wlink(h_i,s_i)=\mathtt{NULL}$) \nextline
          \algline \hspace*{2cm} Insert suffixes $S_i$, $S_{i+1}$, $\ldots$, $S_d$ (Section~\ref{subsec:main-shallow}) \nextline
          \algline \hspace*{2cm} \textbf{return} \nextline
          % \algline  \hspace*{1cm} \textbf{if} ($h_i= h_R$) AND  ($\wlink(h_i,s_{\mathrm{pref}}[1..\ell])=\mathtt{NULL}$) \nextline
          
          % \algline  \hspace*{2cm} \textbf{while} ($\wlink(h_i,s_{\mathrm{pref}}[1..\ell])=\mathtt{NULL}$) AND ($\ell> 0$) \nextline
          % \algline  \hspace*{3cm} $\ell\leftarrow \ell-1$ \nextline
          % \algline  \hspace*{2cm} \textbf{endwhile} \nextline
          % \algline  \hspace*{1cm} \textbf{endif} \nextline
          
          % \algline  \hspace*{1cm} \textbf{if} $h_i\not=\mathtt{NULL}$ \nextline
          \algline \hspace*{1cm} Insert $S_i=s_iS_{\mathrm{init}}$ into the PST \nextline
          %\algline  \hspace*{1cm} \textbf{endif} \nextline
          % \algline  \hspace*{1cm} find the small subtree where $s_{\mathrm{pref}}S_{\mathrm{init}}$ must be inserted \nextline
          \algline \textbf{until} ($i=d$) \nextline
          % \algline  \textbf{if} ($i<d$) \nextline
          % \algline  \hspace*{1cm} locate $s_{\mathrm{pref}}$ in the PST and insert $s_{\mathrm{pref}}S_{\mathrm{init}}$ into the PST \nextline
          % \algline  \hspace*{1cm} $i\leftarrow i+1$ \nextline
          % \algline  \hspace*{1cm} \textbf{while} ($i<d$) \nextline
          % \algline  \hspace*{2cm} $s_{\mathrm{pref}}\leftarrow a_{d-i}s_{\mathrm{pref}}$ \nextline
          % \algline  \hspace*{2cm} Insert $s_{\mathrm{pref}}S_{\mathrm{init}}$ into the PST \nextline
          % \algline  \hspace*{2cm} $i\leftarrow i+1$ \nextline
          % \algline  \hspace*{1cm} \textbf{endwhile} \nextline
          % \algline  \textbf{endif} \nextline
        \end{tabular}
      \end{flushleft}
    \end{minipage}
  }
  \caption{\label{fig:append}Prepending new suffixes. $h_R$ is the path that contains the root node.}
\end{figure}

\paragraph{Insertion of a Suffix $S_i$.}
We now explain how suffix $S_i=a_{d-i+1}\ldots a_dS_{\init}$ for some $i\ge 1$ is inserted. Let $h'_i\not = \mathtt{NULL}$ be the centroid path that is reached by following the w-link from $h_i$ labeled with $a_{d-i+1}\ldots a_d$. Let $l_{i-1}$ be the leaf, known from the previous iteration, that holds $a_{d-i+2}\ldots a_dS_{\init}$ ($l_{i-1}=l_0$ if $i=1$). % We assume that $h'_i\not = \mathtt{NULL}$; the leaf $l_{i-1}$ is known from the previous iteration.

%\yakov{Notation: I denote by the term ``location'' both a location on an edge and a node in the tree. I think it is probably not standard. Leave it as it is? with an explanation? Or may be, change ``location'' to ``position'' or something else?}\\

A \emph{location} in the big subtree of the PST is identified by a centroid path and a string depth. A location in a small subtree is identified by a position on a tree edge. Thus a location can correspond to a tree node or to a position on a tree edge.
We will either find the location $\nu_b$ such that the new leaf $l_i$ holding $S_i$ must be a child of $\nu_b$ or identify the small subtree where $l_i$ must be inserted.
Thus $\nu_b$ is the lowest location such that (1) $\nu_b$ is labeled with a prefix of $S_i$
and (2) $\nu_b$ is either a big node or a location on an edge connecting a big node to some other node. 
% First, we find the lowest location $\nu_b$  that is labeled with a prefix of $s_{\mathrm{pref}}S_{\mathrm{init}}$, such that

We say that a location \(\nu\) matches a string \(s\) if
\(\operatorname{label}(\nu)=s\).

We first locate the lowest matching location within the big
subtree. Afterward, we check whether the match continues along
an edge to a small subtree. In the following lemma, $\nu_l$ denotes the lowest matching location within the big subtree.
The lemma shows that we need to examine $O(1)$ centroid paths in order to find $\nu_b$.
\begin{lemma}
  \label{lemma:ubpath-alt2}
Let \(s_i=a_{d-i+1}\cdots a_d\), \(S_i=s_iS_{\mathrm{init}}\),
and \(h'_i=\operatorname{wlink}(h_i,s_i)\).
Let \(\nu_a\) be the lowest ancestor location of
\(\operatorname{tail}(h'_i)\) that matches a prefix of \(S_i\),
and let $\nu_l$ be the lowest location in the big subtree
that matches a prefix of \(S_i\).
Either (i) $\nu_l$ is an ancestor of $\tail(h'_i)$ and $\nu_l=\nu_a$ or
(ii) $\nu_a$ is a node and $\nu_l$ is a location on a centroid path $h'_p$ such that $\parent(\head(h'_p))=\nu_a$, and $h'_p$ is a leaf of the centroid-path tree.

\commented{
If $\nu_l$ is an ancestor of \(\operatorname{tail}(h'_i)\),
then $\nu_l=\nu_a$. Otherwise, \(\nu_a\) is an explicit node.
Its child \(v\) toward $\nu_l$ is the head of a centroid path
\(h'_p\) that is a child of the path containing \(\nu_a\), and
\(\operatorname{weight}(v)<2\tau\).
Consequently, \(h'_p\) is a leaf in the centroid tree, and
$\nu_l$ lies on \(h'_p\).
}
\end{lemma}

%For w-links, an implicit target location on the incoming edge
%of a big centroid path is associated with that path.
%Throughout the batch, the source paths traversed by the upward
%search retain their identities.

\begin{proof}
If $\nu_l$ is an ancestor of \(\operatorname{tail}(h'_i)\),
then $\nu_l=\nu_a$. Otherwise, the paths to $\nu_l$ and
\(\operatorname{tail}(h'_i)\) diverge at the node
\(\nu_a\). Consider the child \(v\) of $\nu_a$ such that $v$ is the highest node on the path from $\nu_a$ to $\nu_l$ or the highest node below $\nu_l$. The node is big and is the
head of some centroid path \(h'_p\).

Suppose for contradiction that \(\operatorname{weight}(v)\geq 2\tau\).
Let $\rho$ denote the location on the edge $(\nu_a,v)$ with string depth $\strdepth(\nu_a)+1$. 
  There are at least $2\tau$ suffixes that begin with $\slabel(v)$; hence there are $\geq 2\tau$ suffixes that begin with $\slabel(\rho)$.
  Let $s_{\mu}$ be the string obtained by removing the first $i$ symbols from $\slabel(\rho)$ and let $\mu$ be the location with $\slabel(\mu)=s_{\mu}$. Since there are $\geq 2\tau$ suffixes that begin with $\slabel(\rho)$, there are $\geq 2\tau$ suffixes that begin with $\slabel(\mu)$ and the weight of $\mu$ is at least $2\tau$. By our choice of $\rho$ and $\nu_a$, the location $\mu$ is on some path $h_p$ that is a child of $h_i$ and an ancestor of $h_0$: if it were on $h_i$ or on one of its ancestors, then $\rho$ would be on $h'_i$ or on one of its ancestors. But there is a big node $\head(h_p)$ at or below $\mu$. Hence $\wlink(h_p,s_i)\not=\mathtt{NULL}$. This contradicts our choice of $h_i$ as the lowest ancestor of $h_0$ with $\wlink(h_i,s_i)\not=NULL$.
  Thus $\weight(v)<2\tau$. Hence the path $h'_p$ is a $0$-level path and it has no descendants in the centroid-path tree. Since $\nu_l$ is a descendant of $\rho$, $\nu_l$ must be on $h'_p$. 

\end{proof}

\begin{figure}[bt]
  \centering
  % NEW FIGURE for Lemma 2 (old figure: \includegraphics[page=3,width=.4\textwidth]{append1})
  \begin{tikzpicture}[x=1cm,y=1cm,
    nd/.style={circle,fill=black,inner sep=1.3pt},
    loc/.style={rectangle,fill=black,inner sep=1.7pt},
    lab/.style={font=\footnotesize},
    ed/.style={thick},
    zz/.style={thick,decorate,decoration={zigzag,amplitude=1.2pt,segment length=5pt}}]
  % ---------- target side
  \node[nd] (root) at (2.6,0.6) {};
  \node[lab,above] at (root) {root};
  \node[nd] (nua) at (0,-2) {};
  \draw[zz] (root) -- (nua);
  \node[lab,left] at (nua) {$\nu_a$};
  % towards tail(h'_i)
  \node[nd] (ti) at (-1,-4.9) {};
  \draw[zz] (nua) -- (ti);
  \node[lab,left] at (-0.55,-3.5) {$h'_i$};
  \node[lab,below] at (ti) {$\tail(h'_i)$};
  % edge to v = head(h'_p)
  \node[nd] (v) at (1,-3.1) {};
  \draw[ed] (nua) -- (v);
  \node[loc] (rho) at ($(nua)!0.3!(v)$) {};
  \node[lab,below=1pt] at (rho) {$\rho$};
  \node[lab,right] at (v) {$v=\head(h'_p)$};
  \draw[ed,dashed] (v) -- (1.3,-4.9);
  \node[loc] (nul) at ($(v)!0.55!(1.3,-4.9)$) {};
  \node[lab,right] at (nul) {$\nu_l$};
  \node[lab,below] at (1.3,-4.9) {$h'_p$};
  % ---------- source side
  \node[nd] (x) at (5.2,-1.7) {};
  \draw[zz] (root) -- (x);
  \draw[zz] (x) -- (5.0,-3.2);
  \node[lab,left] at (5.05,-2.6) {$h_i$};
  \node[nd] (hp) at (6.2,-2.9) {};
  \draw[ed] (x) -- (hp);
  \node[loc] (mu) at ($(x)!0.55!(hp)$) {};
  \node[lab,above right] at (mu) {$\mu$};
  \node[lab,right] at (hp) {$\head(h_p)$};
  \draw[zz] (hp) -- (6.4,-4.2);
  \node[lab,right] at (6.35,-3.6) {$h_p$};
  \draw[ed,dashed] (6.4,-4.2) -- (6.6,-4.9);
  \node[lab,below] at (6.6,-4.9) {$h_0$};
  % ---------- generalized Weiner link labeled with s_i
  \draw[densely dotted,thick,-{Latex[length=2mm]}] (mu) to[bend right=14] node[lab,above] {$s_i$} (rho);
  \end{tikzpicture}
  \caption{\label{fig:ubpath} Illustration for Lemma~\ref{lemma:ubpath-alt2}. Locations $\mu$ and $\rho$ are shown as squares. Location $\nu_l$ must be on $h'_p$ because $h'_p$ is a leaf path. The dotted arrow is the generalized Weiner link labeled with $s_i$ from $\mu$ to $\rho$.}
\end{figure}
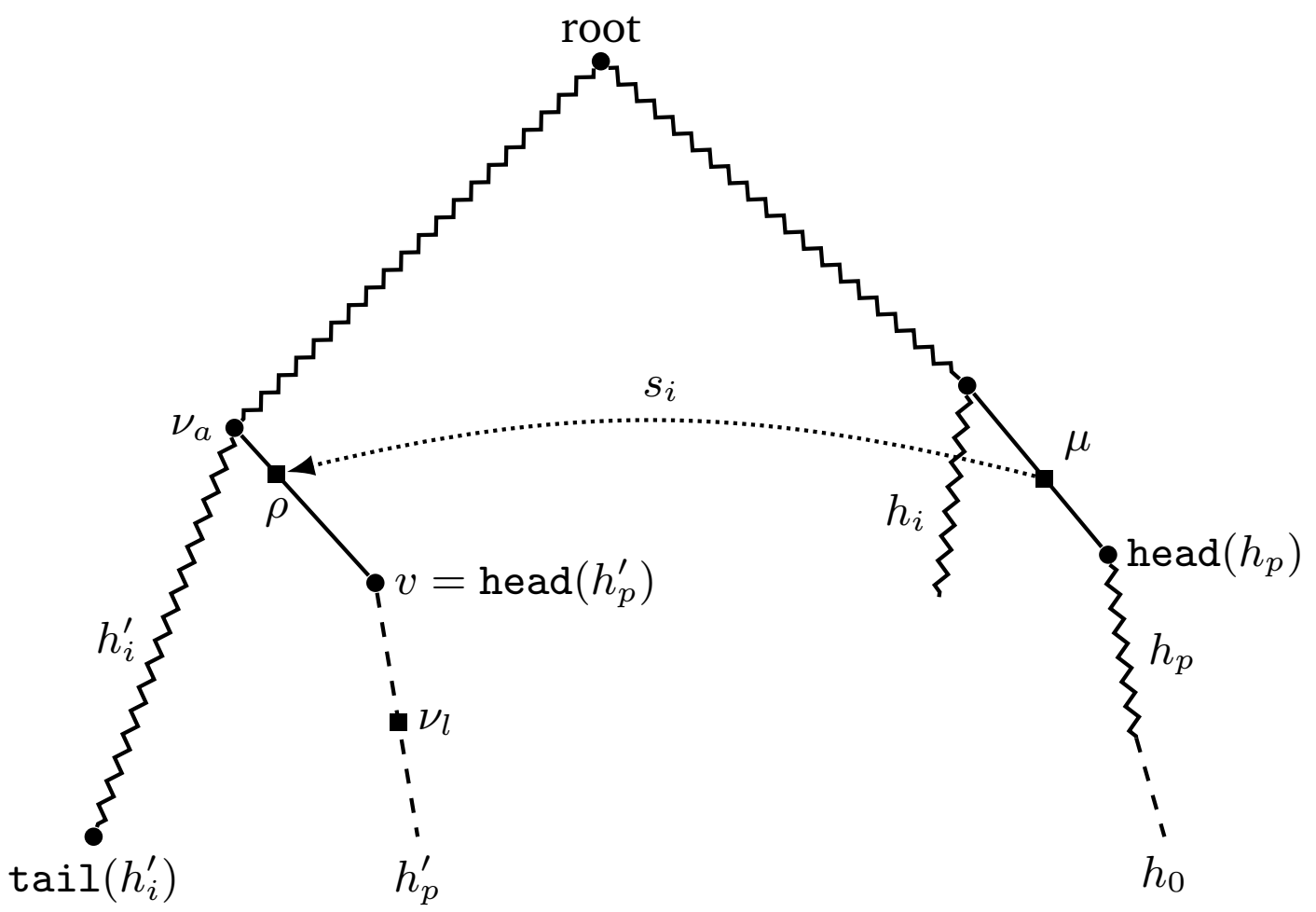

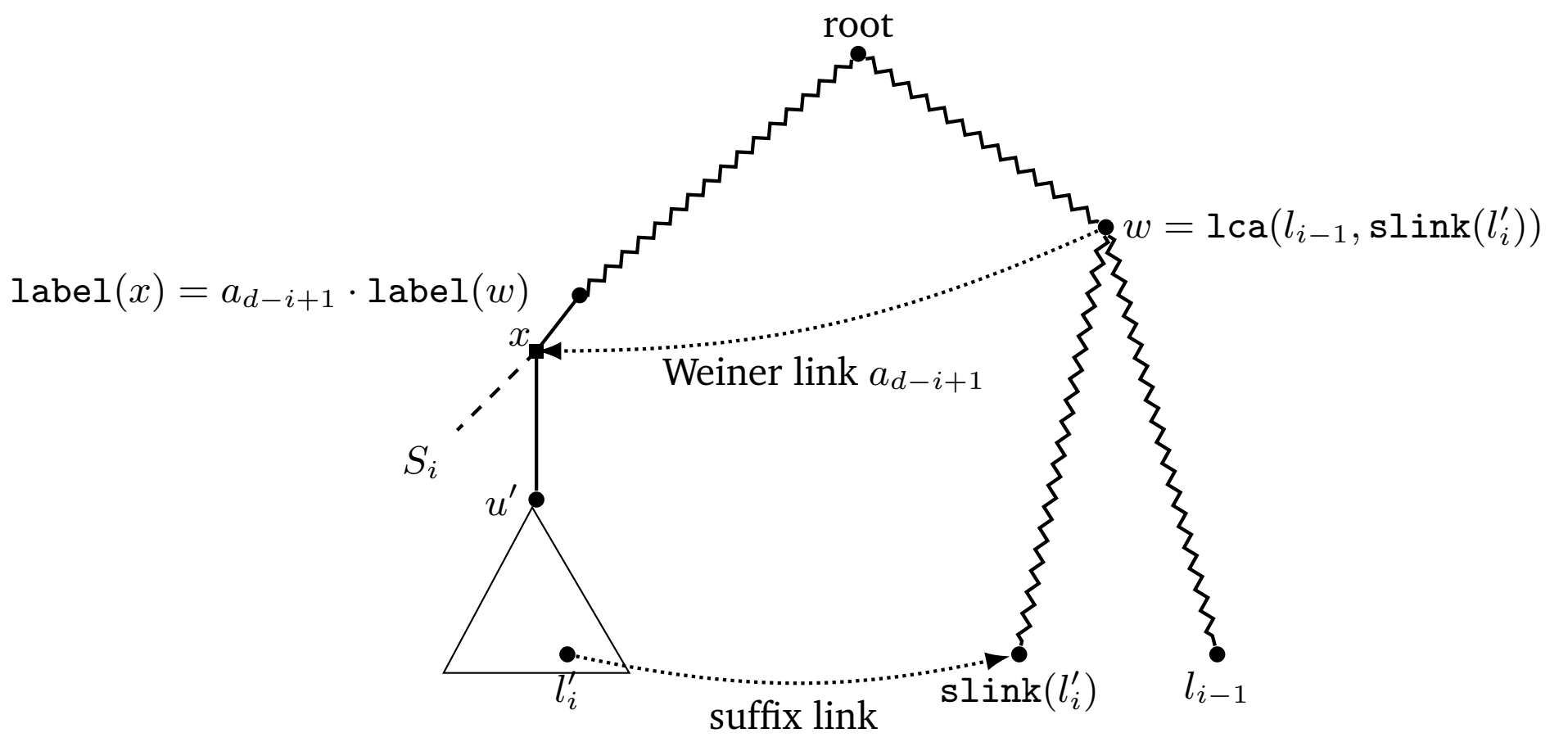
\begin{figure}[bt]
\centering
% NEW FIGURE for Lemma 3 (old figure: \includegraphics[width=.5\textwidth]{append1.pdf})
\begin{tikzpicture}[x=1cm,y=1cm,
  nd/.style={circle,fill=black,inner sep=1.3pt},
  loc/.style={rectangle,fill=black,inner sep=1.6pt},
  lab/.style={font=\footnotesize},
  ed/.style={thick},
  sl/.style={densely dotted,thick,-{Latex[length=2mm]}}]
% root
\node[nd] (r) at (2.6,0) {};
\node[lab,above] at (r) {root};
% ---- left: path labeled a_{d-i+1}..., node u', leaf l'_i
\coordinate (x) at (0,-2.4);
\node[nd] (up) at (0,-3.6) {};
\draw[ed,decorate,decoration={zigzag,amplitude=1.2pt,segment length=5pt}] (r) -- (0.35,-1.95);
\draw[ed] (0.35,-1.95) -- (x) -- (up);
\node[nd] at (0.35,-1.95) {};
\node[loc] at (x) {};
\node[lab,above left,align=right,inner sep=1pt] at (x) {$\lab(x)=a_{d-i+1}\cdot\lab(w)$\\[-1pt]$x$};
\node[lab,left] at (up) {$u'$};
\draw (up) -- (-0.75,-5.0) -- (0.75,-5.0) -- cycle;
\node[nd] (lp) at (0.25,-4.85) {};
\node[lab,below] at (lp) {$l'_i$};
% the new suffix S_i leaves the path at x
\draw[dashed,thick] (x) -- (-0.64,-3.04);
\node[lab,below left] at (-0.64,-3.04) {$S_i$};
% ---- right: w = lca(l_{i-1}, slink(l'_i))
\node[nd] (w) at (4.6,-1.4) {};
\draw[ed,decorate,decoration={zigzag,amplitude=1.2pt,segment length=5pt}] (r) -- (w);
\node[lab,right] at (w) {$w=\lca(l_{i-1},\slink(l'_i))$};
\node[nd] (sl) at (3.9,-4.85) {};
\node[nd] (lm) at (5.5,-4.85) {};
\draw[ed,decorate,decoration={zigzag,amplitude=1.2pt,segment length=5pt}] (w) -- (sl);
\draw[ed,decorate,decoration={zigzag,amplitude=1.2pt,segment length=5pt}] (w) -- (lm);
\node[lab,below] at (sl) {$\slink(l'_i)$};
\node[lab,below] at (lm) {$l_{i-1}$};
% ---- links
\draw[sl] (lp) to[bend right=12] node[lab,below,pos=0.5] {suffix link} (sl);
\draw[sl] (w) to[bend left=12] node[lab,below=3pt,pos=0.5] {Weiner link $a_{d-i+1}$} (x);
\end{tikzpicture}
\caption{\label{fig:append1}Illustration for Lemma~\ref{lemma:anc} in the case $t+1<\strdepth(u')$. The leaf $l_{i-1}$ holds $S_{i-1}$, and $w=\lca(l_{i-1},\slink(l'_i))$ has string depth $t$. The location $x$ with $\lab(x)=a_{d-i+1}\cdot\lab(w)$ has string depth $t+1$ and lies on the path from the root to $l'_i$; the new suffix $S_i=a_{d-i+1}S_{i-1}$ leaves this path at $x$, so $|\lcp(S_i,\lab(u'))|=t+1$.}
\end{figure}

The following lemma computes the length of the longest common prefix of \(S_i\) and the label of a given node; for two strings $x$ and $y$, $\lcp(x,y)$ denotes their longest common prefix. We set $S_0=S_{\init}$.
\begin{lemma}
  \label{lemma:anc}
  % Let $u_b$ denote the lowest node  that is labeled with a prefix of $s_{\mathrm{pref}}S_{\mathrm{init}}$.
  Suppose that $\slabel(u')$ starts with $a_{d-i+1}$ for some internal node $u'$ in the PST. Let $l'_i$ denote an arbitrary leaf descendant of $u'$. Then $|\lcp(S_i,\slabel(u'))|=\min(\strdepth(u'),t+1)$ for $t=\strdepth(\lca(l_{i-1},\slink(l'_i)))$.
\end{lemma}
\begin{proof}
  If the string $\slabel(u')$ is a prefix of $S_i$, then \[|\lcp(S_i,\slabel(u'))|=|\slabel(u')|=\strdepth(u') \]
  and $|\lcp(S_i,\slabel(l'_i))|\geq \strdepth(u')$ for any leaf descendant $l'_i$ of $u'$.
  If $\slabel(u')$ is not a prefix of $S_i$, then
  \[|\lcp(S_i,\slabel(u'))|= |\lcp(S_i,\slabel(l'_i))|< \strdepth(u')\] for any leaf descendant $l'_i$ of $u'$. Therefore
  \[|\lcp(S_i,\slabel(u'))|= \min(|\lcp(S_i,\slabel(l'_i))|, \strdepth(u')) .\]
  It remains to show that $|\lcp(S_i,\slabel(l'_i))|=t+1$; see Fig.~\ref{fig:append1}. 
  Since $S_i=a_{d-i+1}\cdot S_{i-1}=a_{d-i+1}\cdot \slabel(l_{i-1})$ and $\slabel(l_i')=a_{d-i+1}\cdot \slabel(\slink(l'_i))$,
  \[ |\lcp(S_i,\slabel(l'_i))|=|\lcp(\slabel(l_{i-1}),\slabel(\slink(l'_i)))|+1 .\]
  Since
  \[|\lcp(\slabel(l_{i-1}),\slabel(\slink(l'_i)))|=\strdepth(\lca(l_{i-1},\slink(l'_i)))=t ,\]
  $|\lcp(S_i,\slabel(l'_i))|=t+1$. 
\end{proof}

Now we describe the procedure for suffix insertion.
Initially, we set \(u_i=\mathrm{NULL}\).
We apply Lemma~\ref{lemma:anc} to \(\operatorname{tail}(h_i')\) and find the
lowest ancestor location \(\nu\) of this node whose label is a
prefix of \(S_i\), using the path-ancestor structure of Section~\ref{sec:path-ancestor}.
If \(\nu\) is an explicit node \(u\), we check whether \(u\) has
a child \(v\) such that the first symbol on the edge \((u,v)\)
equals \(S_i[\operatorname{strdepth}(u)+1]\).
If such a child exists and is big, then
\(v=\operatorname{head}(h_p')\) for some centroid path \(h_p'\).
We apply Lemma~\ref{lemma:anc} to \(\operatorname{tail}(h_p')\) and replace
\(\nu\) by the lowest ancestor location of this node whose label
is a prefix of \(S_i\).
By Lemma~\ref{lemma:ubpath-alt2}, \(h_p'\) is a leaf in the centroid-path tree.
After this step, any further matching child must be small.
If \(\nu\) is not an explicit node, or if the node at \(\nu\)
has no child matching the next symbol of \(S_i\), we set
\(\nu_b=\nu\).
Otherwise, let \(u\) be the node at \(\nu\) and let \(u_i\)
be its matching small child.
Choose an arbitrary leaf descendant \(l'\) of \(u_i\), taking
\(l'=u_i\) if \(u_i\) is a leaf, and compute
\[
t=\min\left\{
\operatorname{strdepth}(u_i),
1+\operatorname{strdepth}\bigl(
\lca(l_{i-1},\operatorname{slink}(l'))
\bigr)
\right\}.
\]
By the argument of Lemma~\ref{lemma:anc}, \(t\) is the length of the longest
common prefix of \(S_i\) and \(\operatorname{label}(u_i)\).
We set \(\nu_b\) to the location at string depth \(t\) on the
edge \((u,u_i)\) (possibly $\nu_b=u_i$).

Once the location $\nu_b$ is known, we can finish the insertion of the suffix $S_i$. 
If $\nu_b=u_i$, we insert the suffix $S_i$ into the small subtree rooted at $u_i$. The procedure for inserting a suffix into a small subtree is described in Section~\ref{subsec:small-subtr}.
If $\nu_b\not = u_i$ but there is a node $u_b$ at location $\nu_b$, then we create a new child $l_i$ of $u_b$. The dictionary $\child_{u_b}$ is updated accordingly.
Otherwise $\nu_b$ is a location on some edge $(u,v)$. We create a node $u_b$ at this location. Then we create a new child $l_i$ of $u_b$ and update the dictionaries: we insert $l_i$ and $v$ into $\child_{u_b}$ and we insert $u_b$ into $\child_u$; if $v$ is big, we also insert $u_b$ into $\Dict_h$, where $h$ is the centroid path of $v$. In the special case when $\nu_b$ is a location on the edge $(u,u_i)$ leading to a small node, we make $u_b$ the new root of the small subtree $T(u_i)$. After the new leaf $l_i$ is created, we set $\slink(l_i)=l_{i-1}$
and mark $l_{i-1}$ with $a_{d-i+1}$, the symbol of its Weiner
link to $l_i$ (Section~\ref{subsec:small-subtr}). We also insert
$l_i$, and the new internal node $u_b$ if one was created, into
the dynamic lowest-common-ancestor structure
(Section~\ref{sec:path-ancestor}), and we update the weight
counters (details will be given in the full version). These updates take $O(1)$
time. If $l_i$ is inserted into a small subtree, the corresponding
set $R(\cdot)$ is updated as described in
Section~\ref{subsec:small-subtr}. By Lemma~\ref{lemma:insert-no-wlink} in Section~\ref{sec:wlink-nodes}, the insertion of $S_i$ does not change any w-link.

We also need to consider one additional situation. Suppose that our procedure reached the root centroid path $h_R$ and there is no w-link from $h_R$ that is labeled with $a_{d-i+1}\ldots a_d$; see lines 11--13 in Fig.~\ref{fig:append}. In this case the remaining suffixes can be inserted as described in Section~\ref{subsec:main-shallow}.

\paragraph{Running Time.}
We now bound the time needed to insert the suffixes
$S_1,\ldots,S_d$, excluding insertions into small subtrees
(Section~\ref{subsec:small-subtr}) and the insertion of the
remaining suffixes in the case of a missing w-link
(Section~\ref{subsec:main-shallow}).
In the procedure of Fig.~\ref{fig:append}, the path $h_i$
starts at $h_{i-1}$ and only moves upward in the
centroid-path tree. Hence the total number of iterations of
the inner while-loop, over all $i$, is bounded by the height
of the centroid-path tree, which is $O(\log n)$.
Each w-link test takes $O(1)$ time, as explained above.
To insert a single suffix $S_i$, we apply Lemma~\ref{lemma:anc}
$O(1)$ times; each application uses one suffix link and one
lowest-common-ancestor query (Section~\ref{sec:path-ancestor}).
We also perform $O(1)$ queries to the path-ancestor structures
and $O(1)$ lookups and updates in the dictionaries $\Dict_h$
and $\child_u$. All these operations take $O(1)$ time (with
high probability for the dictionary updates). Therefore, the
$d$ suffixes are inserted in $O(\log n+d)=O(d)$ time in total.

\section{Main Text: Adding a Leaf to a Small Subtree}
\label{subsec:small-subtr}
In this section, we explain how a suffix $S_i=a_{d-i+1}a_{d-i+2}\ldots a_dS_{\init}$, $1\le i\le d$, can be inserted into the small subtree $T(v_i)$. 

\paragraph{Data Structure for a Small Subtree.}
% BEGIN EDIT: Item 8.1 (small-subtree notation and insertion)
% OLD: For every small tree $T(v)$ we maintain the set $R(v')$ where $v'$ is the node reached by following the suffix link from $v$. $R(v')$ contains leaves $sl_j=\slink(l_j)$ for all leaf descendants $l_j$ of $v$. We identify a leaf $l_{\max}$ such that $\strdepth(\lca(l_{\max},l')\ge \strdepth(\lca(l,l'))$ for every $l_j\in R(v)$. Once a leaf $l_{\max}$ and $t=\strdepth(\lca(l_{\max},l')$ are known, we can insert $S_i$ into $T(v_i)$.  We find the leaf $l_s$ such that $\slink(l_s)=l_{\max}$ and find the  ancestor $\nu$ of $l_s$ with string depth $t+1$. If there is no node $u$ at locus $\nu$, we insert a new node $u$ at this locus. Then  $l_{n}$ can be added as a new child of $u$.  In the rest of this section we describe how we can find the leaf $l_{\max}$.  
% OLD: 
% NEW:
For every small subtree {$T(v)$}, let {$v'=\slink(v)$} and let {$a$} be
the first symbol of {$\lab(v)$}, so that {$\lab(v)=a\,\lab(v')$}.
We maintain the set {$R(v')$} consisting of the leaves
{$\slink(x)$} for all leaf descendants {$x$} of {$v$}.
The notation {$R(v')$} suppresses its dependence on {$a$} (and hence on
the source subtree {$T(v)$}); sets and their associated data structures
are kept separately for different source subtrees, even when their
suffix-link targets coincide. Each leaf in this set stores, with
its entry in the set, a pointer to the corresponding source leaf {$x$}.

For the insertion of {$S_i$} into {$T(v_i)$}, we set {$v=v_i$},
{$l=l_{i-1}$}, and {$R=R(v')$}. We identify a leaf {$l_{\max}\in R$}
such that {$\strdepth(\lca(l_{\max},l))\geq\strdepth(\lca(x,l))$}
for every {$x\in R$}, and set {$t=\strdepth(\lca(l_{\max},l))$}.
Using the stored pointer, we retrieve the source leaf {$l_s$} satisfying
{$\slink(l_s)=l_{\max}$}. We locate the ancestor location {$\nu$} of
{$l_s$} at string depth {$t+1$}. If {$\nu$} lies inside an edge, we split
that edge by inserting a node {$u$} at {$\nu$}; otherwise, let {$u$} be
the existing node at {$\nu$}. We then add the new leaf {$l_i$} storing
{$S_i$} as a child of {$u$}. In the rest of this section we describe how
to find {$l_{\max}$}.

% END EDIT: Item 8.1 (small-subtree notation and insertion)
\paragraph{List of Leaves.}
For a leaf \(l\), let \(w_l\) denote its lowest big ancestor.
We define \(\spath(l)\) to be the centroid path containing
\(w_l\).
All centroid paths are stored in pre-order in a list $L_C$: Children $h_i$ of the same path $h$ are ordered by the lexicographic order of $\leaf(h_i)$.
% BEGIN EDIT: Item 8.2 (List of Leaves)
% OLD: We also maintain the list $L_L$ that contains all leaf nodes of the PST, positions of leaves in $L_L$ are determined by positions of their centroid paths in $L_C$: let the order of paths in $L_C$ be $L_C=\{\,g_1,g_2,\ldots, gs\,\}$.  $L_L$ All leaves $l$ such that $\spath(l)=g_1$ are stored at the beginning of $L_L$. For any $i$, $1< i\leq s$ , suppose  that  leaves $l_1$, $l_2$, and $l_3$, $l_1$ are such that $\spath(l_1)=g_{i-1}$,  $\spath(l_2)=g_{i}$, and  $\spath(l_3)=g_{i+1}$. Then $l_1$ precedes $l_2$ and $l_2$ precedes $l_3$ in $L_L$. If $\spath(l)=\spath(l')$, then $l$ and $l'$ can appear in an arbitrary order in $L_L$.
% NEW:
We maintain a list {$L_L$} containing all leaves of the PST.
Let {$L_C=(g_1,g_2,\ldots,g_s)$} be the preorder list of centroid
paths. The list {$L_L$} is the concatenation, for {$i=1,\ldots,s$},
of the groups of leaves {$l$} satisfying {$\spath(l)=g_i$}.
Within each group, the leaves may appear in an arbitrary order.
% END EDIT: Item 8.2 (List of Leaves)
We say that a leaf $l$ is smaller (larger) than a leaf $l'$ with respect to $L_L$ iff $l$ appears
before (after) $l'$ in $L_L$. 
\paragraph{Finding $l_{\max}$.}
% BEGIN EDIT: Item 2 (per-path entries)
% OLD: For every set $R(v)$ we maintain the set containing  centroid paths of its leaves
% OLD: \(H(v)=\{\operatorname{path}(l):l\in R(v)\}\).  For every centroid path $h\in H(v)$, we keep a data structure $D(h)$ .  The set $L_H$ contains one representative leaf $l_h$ such that $\spath(l_h)=h$  for every $h\in H(v)$.  We define the predecessor (resp.\ the successor) of a leaf $l_q$ in $L_H$ as the largest (smallest) leaf in $L_H$ that is smaller than or equal to $l_q$ (resp.\ larger than or equal to $l_q$). Here smaller and larger is with respect to $L_L$.
% OLD: We keep a data structure that supports predecessor queries on $L_H$. 
% OLD: Among all leaves $l\in R(v)$ such that $\spath(l)=h$,  we identify a leaf $l$ with the largest string depth of its lowest big ancestor and store it in the dictionary $\Lowest(h)$, $\Lowest(h)=l$.
% NEW:
To simplify the notation, in the rest of this section we write $R(v)$, $H(v)$, $L_H(v)$, and $P_a(v)$ for a set of the form defined above and its associated structures, where $v$ denotes the node below which the leaves of the set lie (the suffix-link target, denoted $v'$ above). For every set {$R=R(v)$} we maintain the set of centroid paths
{\(H(v)=\{\spath(x):x\in R\}\)}.
The set {$L_H$} contains one representative leaf {$l_h\in R$} with
{$\spath(l_h)=h$} for every $h\in H(v)\setminus\{g_R(v)\}$, where $g_R(v)$ is the centroid path of $v$.
We define the predecessor (resp.\ successor) of a leaf {$l_q$} in {$L_H$}
as the largest (resp.\ smallest) leaf in {$L_H$} that is smaller than or
equal to {$l_q$} (resp.\ larger than or equal to {$l_q$}), with respect to {$L_L$}.
We keep a data structure that supports predecessor and successor queries
on {$L_H$}. For every {$h\in H(v)$}, the dictionary entry {$\Lowest(R,h)$}
stores a leaf {$x\in R$} with {$\spath(x)=h$} that maximizes
{$\strdepth(w_x)$}, where {$w_x$} is the lowest big ancestor of {$x$}.
Ties are broken arbitrarily.
% END EDIT: Item 2 (per-path entries)

\begin{lemma}
  \label{lemma:lmax-pred}
  Let \(\spath(l)=h\) and let $h_1$ denote the child of $h$ in the centroid-path tree, such that the string depth of $\parent(\head(h_1))$ is maximized. Let $l_1$ denote an arbitrary leaf descendant of $\head(h_1)$. If $h$ is a leaf path, we set $h_1=h$ and $l_1=l$.
%and let \(l_1\) be a leaf descendant of \(\operatorname{tail}(h)\) such that \(\spath(l_1)\) is a proper descendant of \(h\).
  Let \(l_p\) and \(l_s\) be the predecessor and successor of \(l_1\) in \(L_H\), and write \(h_p=\spath(l_p)\) and \(h_s=\spath(l_s)\). Then an optimal leaf \(l_{\max}\), i.e., a leaf $x\in R(v)$ maximizing $\strdepth(\lca(x,l))$, can be chosen such that
\[
\spath(l_{\max})\in
\left\{\,h,\ h_p,\ h_s,\
\operatorname{LCA}(h,h_p),\
\operatorname{LCA}(h,h_s),\
g_R(v)\
\right\}.
\]
Here, LCAs of paths are taken in the centroid-path tree,
and terms involving nonexistent neighbors are omitted.
\end{lemma}
\begin{proof}
Let \(l_{\max}\in R(v)\) be an optimal leaf, and write
\(h_{\max}=\spath(l_{\max})\), \(z=\lca(l,l_{\max})\).
Any leaf of \(R(v)\) below \(z\) is also optimal.
If \(h_{\max}=h\), the conclusion follows because \(h\)
is a candidate. Hence assume \(h_{\max}\neq h\), and let
\(h_z=\operatorname{LCA}(h,h_{\max}) \)
in the centroid-path tree. Then \(z\) is a big node on \(h_z\).

Consider the child paths of \(h_z\) whose heads lie below
\(z\). These children are consecutive in the ordering by
$\leaf(h')$ that is used in the definition of $L_C$: their subtrees lie in the
lexicographic interval of leaves below \(z\), whereas every
other child subtree lies outside that interval.
Consequently, these child paths and all their descendants
form a consecutive interval \(I_z\) in \(L_C\).
Every leaf assigned to a path in \(I_z\) lies below \(z\).

\medskip
\noindent
\textbf{Case 1: \(h_{\max}\neq h_z\).}
Then \(h_{\max}\in I_z\). We also have
\(\spath(l_1)\in I_z\).
Indeed, if \(h\neq h_z\), then \(h\in I_z\), and
\(\spath(l_1)\) is either \(h\) or a descendant
of \(h\). If \(h=h_z\), the child path toward \(h_{\max}\)
attaches to \(h\) at \(z\) or below it.
By the choice of \(h_1\), its attachment point is also
at \(z\) or below it. Thus \(h_1\), and hence
\(\spath(l_1)\), belongs to \(I_z\).

The path groups belonging to \(I_z\) form a consecutive
interval in \(L_L\). This interval contains \(l_1\) and
the representative of \(h_{\max}\). Therefore at least
one of \(l_p,l_s\) belongs to this interval.
That leaf lies below \(z\), so it is optimal, and its
path is one of the candidates \(h_p,h_s\).

\medskip
\noindent
\textbf{Case 2: \(h_{\max}=h_z\).}
Here \(h_z\) is a proper ancestor of \(h\), so both \(h\)
and \(\spath(l_1)\) belong to \(I_z\).
If $h_z=g_R(v)$, then $h_{\max}=g_R(v)$ is a candidate, and we are done. Otherwise, $h_z$ has a representative in $L_H$.
The representative of \(h_z\) precedes \(l_1\);
hence \(l_p\) exists.

If \(h_p\in I_z\), then \(l_p\) lies below \(z\) and is
optimal. Otherwise, since \(l_p\) lies between the
representative of \(h_z\) and \(l_1\), its path is either
\(h_z\) itself or lies in an earlier child subtree of
\(h_z\) outside \(I_z\). In either situation,
\[
  \operatorname{LCA}(h,h_p)=h_z=h_{\max}.
\]
Thus the candidate set contains the path of \(l_{\max}\).

In both cases, a candidate path contains an optimal leaf.
\end{proof}

Once the predecessor and successor of {$l_1$} in {$L_H$} are known,
Lemma~{\ref{lemma:lmax-pred}} gives a constant-size set of candidate paths.
For each candidate path {$h'\in H(v)$}, we retrieve {$\Lowest(R,h')$};
candidate paths not represented in {$R$} are ignored.
For leaves assigned to a fixed path and lying outside the small subtree
containing {$l$}, their LCP with {$l$} is nondecreasing with the depth of
their lowest big ancestor: moving the attachment point downward along
the path cannot move its LCA with {$l$} upward.
Thus a deepest attachment suffices unless an optimal leaf lies in the
same small subtree as {$l$}.

To handle this case, let {$T(v_s)$} be the small subtree containing {$l$},
and let {$a$} be the symbol marking the leaves of {$R$}.
Using {$D(v_s)$} (described in the paragraph ``Data Structure for Small Subtrees'' in Section~\ref{subsec:main-shallow}),
we find the inclusive predecessor and successor of {$l$} among the leaves
marked with {$a$} in {$T(v_s)$}, in lexicographic order.
We retain each returned leaf only if it belongs to {$R$}.
For the insertion into $T(v_i)$, write $v'_i=\slink(v_i)$,
so that $R=R(v'_i)$ consists exactly of the $a$-marked leaves below $v'_i$, for the
symbol {$a$} removed by this suffix link,
and the query leaf $l$ is also below $v'_i$.
Consequently, a returned {$a$}-marked leaf {$x$} belongs to {$R$} if and only if
\[
  \strdepth(\lca(l,x))\geq \strdepth(v'_i).
\]
This is a constant-time membership test.
The leaves below $v'_i$ form a lexicographic interval containing $l$.
Hence, whenever {$R\cap T(v_s)$} is nonempty, the two local neighbors
include a leaf of this intersection with maximum LCP with {$l$}:
on either side containing a leaf of {$R\cap T(v_s)$}, the nearest
{$a$}-marked leaf also belongs to {$R$}.
We compare the LCPs with {$l$} of all retrieved dictionary leaves and
retained local neighbors, and return a maximizer as {$l_{\max}$}.
There are only {$O(1)$} candidates, so this computation takes {$O(1)$} time
once the predecessor and successor of {$l_1$} in {$L_H$} are known.
% END EDIT: Item 2 (candidate evaluation and local search)
In the rest of this section, we describe a slow and a fast data structure for predecessor queries on $L_H$.

\paragraph{Slow Predecessor Data Structure.}
Using an order-maintenance data structure~\cite{DietzS87}, we can determine relative positions of any two elements in $L_L$ in $O(1)$ time. Hence we can answer predecessor queries using binary search.  
\begin{lemma}
  \label{l:Rslow}
  There exists a data structure that supports predecessor queries on $L_H$ for any set $R(v)$ in $O(\log \log n)$ time and insertions in $O(\log\log n)$ time.
\end{lemma}
\begin{proof}
  We store all representative leaves of $H(v)$ in a balanced binary search tree, such as a weight-balanced B-tree~\cite{ArgeV03}.  
\end{proof}

\paragraph{Labeling of $L_L$.}
The list \(L_L\) is divided into blocks of size \(\Theta(\log^5 n)\).
Every block is assigned an integer label, so that labels increase
with the order of blocks in \(L_L\). We use the labeling scheme
from~\cite{Willard92}. Whenever a new block is inserted into \(L_L\), the labels
of \(O(\log^2 n)\) other blocks may change.

%We also maintain an order-maintenance data structure for \(L_C\).
%Given two centroid paths \(h_1\) and \(h_2\), we can determine in
%\(O(1)\) time whether \(h_1\) precedes \(h_2\).
%Predecessor and successor queries on \(H(v)\) use this order;
%the query path need not belong to \(H(v)\).

\paragraph{Significant Symbols.}
Every leaf is marked by the symbol of its Weiner link, if such a link
exists. Thus a leaf holding the suffix \(T[i..n]\), with \(i>1\),
is marked by \(T[i-1]\).
Let $\rho=\log^2 n$.
A symbol \(a\) is significant in a node \(u\)
if at least a \(1/\rho\)-fraction of the leaf descendants of \(u\)
are marked with \(a\). A symbol \(a\) is significant in a leaf \(l\)
if it is significant in \(\operatorname{head}(h)\) for some path \(h\)
that is an ancestor of \(\spath(l)\) in the centroid-path tree,
including \(\spath(l)\) itself.
A leaf is significant if its marking symbol is significant in it.

\begin{lemma}
  \label{lemma:block-sign}
% BEGIN EDIT: Item 4 (lemma statement only)
% OLD:   In any block $B$, significant leaves are marked with $O( )$ distinct symbols.
% NEW:
  In any block {$B$}, significant leaves are marked with {$O(\log^3 n)$} distinct symbols.
% END EDIT: Item 4 (lemma statement only)
\end{lemma}
\begin{proof}
Let $h_f$ and $h_e$ be the paths of the first and last leaf of $B$. The paths whose
groups intersect $B$ form an interval of the preorder of the centroid-path tree, and
the subtree of every path $h$ also forms an interval of this preorder, starting
at $h$. Consider a symbol $a$ that is significant at $\head(h)$, where $h$
is an ancestor of $\spath(l)$ for some leaf $l \in B$.
If the subtree of $h$ contains $h_f$ or $h_e$, then $h$ is an ancestor of $h_f$ or $h_e$;
there are $O(\log n)$ such paths, and each head has at most $\rho$ significant
symbols. This gives $O(\rho\log n)$ symbols.
Otherwise the subtree of $h$ lies strictly between $h_f$ and $h_e$, so all leaf
descendants of $\head(h)$ belong to $B$. Then at least
$\weight(\head(h))/\rho \ge \tau/\rho$ leaves of $B$ are marked with
$a$. Since every leaf has one mark, there are at most $|B|\rho/\tau = O(\rho)$ such
symbols. Hence the total number is $O(\rho\log n) = O(\log^3 n)$.
\end{proof}
% \begin{proof}
% For the lowest common ancestor $z$ of all leaves in a block, there are at most $\rho\cdot \log n$ symbols that are significant. There are at most $\frac{\log^5n}{\rho}$ symbols that are significant for some leaves in the block, but are not significant for $z$. Hence the total number of significant symbols is $O(\rho\cdot \log n+\frac{\log^5n}{\rho})=O(\log^3 n)$. 
% \end{proof}

\paragraph{Fast Predecessor Data Structure.}
% BEGIN EDIT: Item 5 (fast block search)
% OLD: All suffixes in $R(v)$ are marked with the same symbol $a$. If $a$ is significant in $\head(\spath(v)))$, then we store a fast predecessor data structure $P_a(v)$ for $R(v)$.  For each block $B$, such that some leaves from $R(v)$ are in $B$ and $a$ is significant in $B$, we store the label of $B$ in $P_a(v)$.  $P_a(v)$ is implemented as a fusion tree. Since there are $O(\log^5 n)$ leaves in $R(v)$, $P_a(v)$ contains $O(\log^5 n)$ different block labels. Hence both queries and updates on $P_a(v)$ are supported in $O(1)$ time.  When we identified the block $B_l$ that contains the predecessor of the query leaf $l$, we can report the maximum element in 
% OLD: $B$ (provided that $l$ is not in $B$) or find the predecessor of $l$ in $B$ (if $l$ is in $B$).
% OLD: In both cases the search can be completed using a variant of the data structure $D(v)$, described
% OLD: on page~\pageref{par:datasmall}
% NEW:
Fix a set {$R=R(v)$} whose leaves are marked with {$a$}, and let
{$L_H=L_H(v)$} be its representative list. We call {$a$} significant in a
block {$B$} if {$B$} contains an {$a$}-marked leaf that is significant.
Thus Lemma~{\ref{lemma:block-sign}} bounds the number of significant
symbols in a block. The fast search below is used when {$a$} is
significant in every block intersecting {$R$}, as required by the query
procedure below.

For each block {$B$}, we keep a variant {$D_B$} of the structure described
in Section~\ref{subsec:main-shallow}. It stores all leaves of {$B$} in their
{$L_L$} order. In {$D_B$}, a leaf is marked with {$a$} only if it is a chosen
representative in some list {$L_H(v)$} associated with {$a$}; other leaves
are unmarked. These auxiliary marks are separate from the original
marks on suffix leaves. Each representative stores a pointer to its
list. Hence {$D_B$} supports inclusive predecessor and successor queries
among the {$a$}-marked representatives of {$B$} in {$O(1)$} time, and we can
check in {$O(1)$} time whether a returned representative belongs to the
queried list {$L_H$}. We also store pointers to the first and last
representatives of each list.
For a fixed $a$, the lists $L_H(v)$ occupy disjoint intervals of $L_L$. Indeed,
every path in $H(v) \setminus \{g_R(v)\}$ lies in the subtree of a child path of
$g_R(v)$ whose head is below $v$; as in the proof of Lemma~\ref{lemma:lmax-pred}, these child paths are
consecutive in $L_C$, so their groups form an interval $J(v)$ of $L_L$. If
$R(v_1')$ and $R(v_2')$ are associated with the same symbol $a$, then $v_1'$ and
$v_2'$ are not ancestors of each other, since otherwise $T(v_1)$ and $T(v_2)$
would be nested; hence $J(v_1)$ and $J(v_2)$ are disjoint. Consequently, every
$a$-marked representative between the endpoints of $L_H$ belongs to $L_H$.

If {$L_H$} meets more than one block, we keep a fusion tree {$P_a(v)$}
containing the labels of precisely those blocks {$B$} for which
{$L_H\cap B$} is nonempty and {$a$} is significant in {$B$}.
Each entry points to its block. Since {$|L_H|\leq |R|=O(\log^5 n)$},
queries and individual updates in {$P_a(v)$} take {$O(1)$} time.
Under the fast-search condition, every block meeting {$L_H$} is stored.
If {$L_H$} is contained in one block, we store a pointer to that block
and use {$D_B$} directly; no structure {$P_a(v)$} is needed.

Consider an inclusive predecessor query for a leaf {$l_q$} in {$L_L$}.
If {$l_q$} precedes the first representative of {$L_H$}, there is no
predecessor; if {$l_q$} is at or after its last representative, we return
that last representative. Otherwise let {$B_q$} be the block containing
{$l_q$}. For a list meeting several blocks, we find in {$P_a(v)$} the
largest stored block label not exceeding the label of {$B_q$}.
If this is {$B_q$}, we use {$D_{B_q}$} to find the inclusive {$a$}-marked
predecessor of {$l_q$}. We return it if it belongs to {$L_H$}; if no such
representative is returned, we move to the preceding stored block.
If the selected block precedes {$B_q$}, or after this fallback, we
return its largest {$a$}-marked representative, found by querying {$D_B$}
at the last leaf of the block. This representative belongs to {$L_H$}:
the selected block contains a representative of {$L_H$}, and, after
the endpoint checks, its right boundary precedes the last
representative of {$L_H$}. The interval property therefore excludes
representatives of another list with symbol {$a$} after those of {$L_H$}
in this block. If there is no preceding stored block, no predecessor
exists. For a list contained in one block, the same endpoint checks
leave only a local query in {$D_B$}.
Successor queries are symmetric, using the following stored block
and its smallest relevant representative when necessary.
Thus each search uses {$O(1)$} fusion-tree and local queries and takes
{$O(1)$} time.

% END EDIT: Item 5 (fast block search)
\paragraph{Queries: Putting All Pieces Together.}
% BEGIN EDIT: Item 3 (query procedure)
% OLD: We must insert suffixes $S_i=a_{d-i+1}\ldots a_dS_{\init}$ for all $i$, $1\le i\le d$, into the PST.
% OLD: For every index $i$, we may have to insert $S_i$ into the small subtree. Every such insertion involves  finding $l_{\max}$. In order to find $l_{\max}$ we have to find the predecessor of $l_{i-1}$ in $R(v_{i-1})$. If $a_{d-i+1}$ is a significant symbol in $R(v_{i-1})$, we use the data structure $P_a(v_{i-1})$ and find the predecessor in $O(1)$ time. Otherwise we use the binary search tree and answer queries in $O(\log \log n)$ time using Lemma~\ref{l:Rslow}.  
% OLD: 
% OLD: 
% OLD: In order to insert a suffix $S_i=a_{d-i+1}\ldots a_dS_{\init}$ into the small tree $T(v_i)$, we use the data structure for $R(v'_i)$ and find $l_{\max}$.  The computation of $l_{\max}$ can involve answering a predecessor query on $R_L(v'_i)$.  the data structure of Lemma~\ref{l:Rfast}
% OLD: 
% OLD: We answer all predecessor queries in $O(\log n)$ time using the following approach. 
% OLD: Suppose that  we must compute $l_p$ and $l_s$ using  $R_L(v'_i)$. If the symbol $a_{d-i+1}$ is significant in all blocks that intersect $R_L(v'_i)$,  we use   the data structure of Lemma~\ref{l:Rfast}. Otherwise we use the data structure of Lemma~\ref{l:Rslow}
% NEW:
Suppose that {$S_i=a_{d-i+1}\ldots a_dS_{\init}$} must be inserted
into the small subtree {$T(v_i)$}. Let {$v'_i=\slink(v_i)$},
{$R=R(v'_i)$}, {$a=a_{d-i+1}$}, and {$l=l_{i-1}$}.
We find a leaf {$l_{\max}\in R$} maximizing its LCP with {$l$} as follows.

Set {$h=\spath(l)$}. If {$h$} is not a leaf of the centroid-path tree,
choose a child path {$h_1$} maximizing
{$\strdepth(\parent(\head(h_1)))$} and set {$l_q=\leaf(h_1)$}.
Otherwise set {$l_q=l$}.
Here {$l_q$} is the auxiliary leaf denoted by {$l_1$} in
Lemma~{\ref{lemma:lmax-pred}}.
For every nonleaf path {$h$}, we store a pointer to such a child {$h_1$};
together with the stored leaf {$\leaf(h_1)$}, this lets us obtain {$l_q$}
in {$O(1)$} time. We also equip the centroid-path tree with a data structure
supporting lowest common ancestor queries in {$O(1)$} time.

We find the predecessor {$l_p$} and successor {$l_s$} of {$l_q$}
in the representative list {$L_H$} associated with {$R$}, ordered by {$L_L$}.
If {$a$} is significant in every block intersecting {$R$}, we use the
fast predecessor structure described in the preceding paragraph;
the successor query is symmetric. Otherwise we use the balanced
search tree of Lemma~{\ref{l:Rslow}}.
Thus a pair of neighbor queries takes {$O(1)$} time in the fast case
and {$O(\log\log n)$} time in the slow case.

Let {$h_p=\spath(l_p)$} and {$h_s=\spath(l_s)$} when these neighbors exist.
By Lemma~{\ref{lemma:lmax-pred}}, we need only consider the paths
\[
  \{h,h_p,h_s,\operatorname{LCA}(h,h_p),
    \operatorname{LCA}(h,h_s), g_R(v'_i) \}.
\]
Terms involving a missing neighbor are omitted, and LCA is taken in
the centroid-path tree. We evaluate these paths using {$\Lowest(R,\cdot)$}
and the local marked-leaf search described after the proof of
Lemma~{\ref{lemma:lmax-pred}}, retaining only leaves belonging to {$R$}.
The leaf with the largest LCP with {$l$} is returned as {$l_{\max}$}.
This evaluation takes {$O(1)$} time after the two neighbors are known.

After $l_i$ has been added to $T(v_i)$, the leaf $l_{i-1} = \slink(l_i)$
belongs to $R(v_i')$. Let $h = \spath(l_{i-1})$ and let $w$ be the lowest
big ancestor of $l_{i-1}$. If $h \in H(v'_i)$, we set $\Lowest(R,h) = l_{i-1}$
when $w$ is deeper than the lowest big ancestor of the current entry. Otherwise
we add $h$ to $H(v'_i)$ and set $\Lowest(R,h) = l_{i-1}$; if $h \neq g_R(v'_i)$,
we also insert $l_{i-1}$ into $L_H$: we mark it in $D_B$ for the block $B$ that
contains it, and add the label of $B$ to $P_a(v'_i)$ if necessary. If $a$ is not
significant in every block intersecting $R$, we also insert $l_{i-1}$ into the
search tree of Lemma~\ref{l:Rslow}; this happens only in steps that use the slow search.
Finally, $l_{i-1}$ receives the mark $a$ in the data structure $D(\cdot)$ of its
small subtree.

It remains to bound the number of insertions that require the slow
neighbor search.
% END EDIT: Item 3 (query procedure)
\begin{lemma}
  Our method uses the data structure from Lemma~\ref{l:Rslow} $O(\log n/\log\log n)$ times during one transferring round.
\end{lemma}
\begin{proof}
Let {$k$} be the largest index such that
{$\wlink(h_k,a_{d-k+1}\cdots a_d)\neq\mathtt{NULL}$};
set {$k=0$} if no such index exists. If {$k=0$}, there are no steps to
analyze. Otherwise set {$\mu_0=\head(h_k)$}. For {$1\leq i\leq k$},
let {$\mu_i$} be the location matching
{$a_{d-i+1}\cdots a_d\,\lab(\mu_0)$}.
These locations exist because the length-{$k$} w-link is non-NULL;
no rounding to explicit nodes is needed.
Let {$u_i$} be the head of the centroid path containing {$\mu_i$}.

For an insertion into {$T(v_i)$}, the source location {$\nu$} of the
w-link leading to {$v_i$} is a descendant of {$\mu_0$}.
Prepending {$a_{d-i+1}\cdots a_d$} to both labels shows that {$v_i$}
is a descendant of {$\mu_i$}. Removing the first symbol shows that
{$v'_i=\slink(v_i)$} is a descendant of {$\mu_{i-1}$}.

Fix {$i$}, {$1\leq i\leq k$}, and write {$a=a_{d-i+1}$}.
If at least a {$9/(4\rho)$}-fraction of the leaves below {$\mu_{i-1}$}
are marked with {$a$}, then at least a {$1/\rho$}-fraction of the leaves
below {$u_{i-1}$} are marked with {$a$}:
{$\weight(u_{i-1})\leq \frac94\weight(\mu_{i-1})$}
because these locations lie on the same centroid path, whose maximum and minimum permitted weights differ by a factor of {$9/4$}.
Thus {$a$} is significant at {$u_{i-1}$} and in every leaf of {$R(v'_i)$}.
Every block intersecting {$R(v'_i)$} therefore contains a significant
leaf marked with {$a$}, so the fast-search condition holds.

Consequently, whenever the slow search is used, fewer than a
{$9/(4\rho)$}-fraction of the leaves below {$\mu_{i-1}$} are marked with {$a$}.
The weight of {$\mu_i$} equals the number of these {$a$}-marked leaves,
and hence {$\weight(\mu_i)<(9/(4\rho))\weight(\mu_{i-1})$}.
Weights do not increase at the other steps. Since
{$\weight(\mu_0)\leq n$} and {$\weight(\mu_k)\geq 1$}, the number of
slow searches is at most {$\log n/\log(4\rho/9)$}.
For {$\rho=\log^2 n$}, this is {$O(\log n/\log\log n)$}.
% END EDIT: Item 6 (weight-decrease proof)
\end{proof}
% BEGIN EDIT: Item 6 (total query time)
% OLD: Hence the total time needed to answer all predecessor queries is $O(\log n)$. 
% NEW:
The total time for these queries is
{$O(k+(\log n/\log\log n)\log\log n)=O(k+\log n)=O(\log n)$},
since {$k\leq d=O(\log n)$}. Insertions into the search tree of Lemma~\ref{l:Rslow} occur only in steps that use the
slow search, so their total time is also $O(\log n)$.
% END EDIT: Item 6 (total query time)

\paragraph{Maintaining the List $L_L$ and the Structures $P_a(v)$.}
% BEGIN EDIT: Item 7 (gradual block rebuilding)
% OLD: We must maintain the invariant that every block $B$ of $L_L$ contains $O(\tau)$ leaves. When the number of leaves in a block equals to $\frac{3}{2}\tau$, we start the process of splitting $B$ into $B$ and $B'$. We create the new block $B''$ and move $\tau$ rightmost leaves from $B$ to $B'$. The process of splitting is distributed over $\tau/4$ following insertions. 
% NEW:
We must maintain the invariant that every block $B$ of $L_L$ contains $O(\tau)$ leaves.
When a block {$B$} reaches size {$3\tau/2$}, we start constructing
two replacement blocks {$B'$} and {$B''$} corresponding to two
consecutive, nearly equal parts of {$B$}. We retain {$B$} and continue
answering queries using its existing labels and data structure
{$D_B$} throughout the rebuilding process. The leaves of {$B$} are
copied gradually into {$B'$} and {$B''$}, together with the construction
of {$D_{B'}$} and {$D_{B''}$}. Every new leaf inserted into {$B$} is also
inserted into the appropriate replacement block.

We distribute the rebuilding work over the next {$\tau/4$}
insertions into {$B$}. Since rebuilding starts when {$B$} contains
{$3\tau/2$} leaves, its size never exceeds
{$3\tau/2+\tau/4=7\tau/4$} during this process. The total rebuilding
work is {$O(\tau)$}, including the updates to the fast predecessor
structures counted below, and is performed at a constant rate
per insertion into {$B$}. Once the replacement blocks and their
associated structures are ready, we retire {$B$} and switch queries
to {$B'$} and {$B''$}.
% END EDIT: Item 7 (gradual block rebuilding)

Additionally, we must update the labeling of blocks: addition of a new block changes the labels of $O(\log^2 n)$ blocks. The number of significant symbols in each block is $O(\log^3 n)$. We maintain \(P_a(v)\) only when \(R(v)\) spans more than one
block. For a fixed symbol \(a\), the lists \(L_H(v)\) occupy
disjoint intervals of $L_L$. Any such interval
that intersects a block \(B\) and another block must cross
one of the two boundaries of \(B\). Hence at most two
structures \(P_a(v)\) contain the label of \(B\) for each
symbol \(a\). Since \(B\) has \(O(\log^3 n)\) significant
symbols, changing its label requires updating
\(O(\log^3 n)\) structures $P_a(v)$. Thus the overall number of updates in fast predecessor data structures is $O(\log^5 n)$.
Since an update on a data structure $P_a(v)$ takes $O(1)$ time, all updates incurred by changing labels after splitting a block take $O(\log^5 n)=O(\tau)$ time. These updates are performed during the rebuilding of $B$, as part of its $O(\tau)$ budget, before queries are switched to $B'$ and $B''$. When the labels of a range of blocks change, we update the blocks one at a time: blocks whose labels increase are processed from right to left, and blocks whose labels decrease are processed from left to right. Hence the labels stored in every structure $P_a(v)$ are increasing at all times, and queries remain correct while the labels are being updated. Relabelings of overlapping ranges are performed one after another.

\section{Main Text: Inserting Suffixes in the Case of a Missing w-Link}
\label{subsec:main-shallow}
In this section, we consider the situation when $h_i=h_R$ and $\wlink(h_i,s_i)=\mathtt{NULL}$. If $\wlink(h_R,s_i)=\mathtt{NULL}$, then there is no location in the big subtree that is labeled with $a_{d-i+1}\ldots a_d$. Hence $a_{d-i+1}\ldots a_d$ occurs fewer than $2\tau$ times in the main text. We show how to find, for all $j$, $i\le j\le d$, the lowest location that is labeled with a prefix of $S_j=a_{d-j+1}\ldots a_dS_{\init}$. Our procedure separately searches for Weiner links in small subtrees and on leaf centroid paths. If neither case applies, we show that the longest
match with an existing suffix has string depth at most \(d\).
We then find the insertion locations of all remaining suffixes in \(O(d)\) total time.

\newcommand{\cL}{{\cal L}}
\newcommand{\cS}{{\cal S}}

Let $\cL$ denote the set of labels for all nodes with weight at least $2\tau$. Let $\cS$ denote the set of all suffixes in the main text $T$. Let $A=a_1a_2\cdots a_d$ denote the block of prepended symbols; thus $s_i=A[d-i+1..d]$ and $S_i=A[d-i+1..d]S_{\init}$. For a string $X$ and a set $Y$ of strings, $\lcp(X,Y)$ denotes the longest among the common prefixes of $X$ and the strings in $Y$. We maintain the invariant that every big node of weight less than \(2\tau\) lies on a leaf path of the centroid-path tree. Consequently, every node on a nonleaf centroid path has weight at least \(2\tau\), and its label belongs to \(\mathcal L\).

% BEGIN EDIT: Section 8, item 3 (Lemmas 8-10 and Corollaries 1-2 replaced by one lemma)
\begin{lemma}
  \label{lemma:lcp-short}
  For $1\le i\le j\le d$:
  (a) if $|\lcp(S_i,\cL)|< i$, then $|\lcp(S_j,\cL)|< j$;
  (b) if $|\lcp(S_i,\cS)|< i$, then $|\lcp(S_j,\cS)|< j$.
\end{lemma}
\begin{proof}
  Let $s_k=A[d-k+1..d]$ denote the prefix of $S_k$ of length $k$; thus $s_i$ is
  a suffix of $s_j$. A string $X$ is a prefix of a label in $\cL$ if and only if
  $X$ occurs at least $2\tau$ times in $T$, and $|\lcp(S_k,\cS)|\ge k$ if and only
  if $s_k$ occurs in $T$. Every occurrence of $s_j$ contains an occurrence of
  $s_i$. Hence, if $s_i$ occurs fewer than $2\tau$ times (resp.\ does not occur),
  then so does $s_j$.
\end{proof}
% END EDIT: Section 8, item 3

\paragraph{Data Structure for Small Subtrees.}
%Small Trees.  All leaves in a small tree marked by a symbol $a$ are stored in a data structure. We can find the closest preceding and following leaves in $O(1)$ time. Insertions are supported in $O(1)$ time.
\newcommand{\Weiner}{\mathtt{Weiner}}
\label{par:datasmall}
For every small subtree $T(v)$, we keep all leaves in a data structure $D(v)$. For each leaf $l$ in $T(v)$ and for any symbol $a$, we can find the preceding or the following leaf marked with $a$ in $O(1)$ time. $D(v)$ also supports insertions in $O(1)$ time. $D(v)$ is organized as follows.
We store all leaves of $D(v)$ in a weight-balanced B-tree $\cT$~\cite{ArgeV03} with parameters $r=(\log n)/2$ and $k=1$. For every internal node $\alpha$ of $\cT$, we maintain a dictionary $\cE(\alpha)$. $\cE(\alpha)$ contains all symbols $a$ such that at least one leaf descendant $l$ of $\alpha$ is marked with $a$. Children $\alpha'$ of $\alpha$ are assigned integer labels in the left-to-right order. When a node $\alpha$ is initially inserted into the tree, we assign the $i$-th child of $\alpha$ the label $i\cdot n^2$. The leftmost and the rightmost sentinel labels of $\alpha$ are set to $0$ and $4rn^2$ respectively.
For each symbol $a$ in $\cE(\alpha)$ we store the predecessor data structure $\cP_a(\alpha)$. $\cP_a(\alpha)$ contains the labels of all children $\alpha'$ of $\alpha$ that have leaf descendants marked with $a$. This data structure is implemented as a fusion tree~\cite{FredmanW93,PatrascuT14} so that predecessor and successor queries as well as insertions are supported in $O(1)$ time.

A query to $D(v)$ is answered as follows. Suppose that we want to find, for a leaf $l$ and for a symbol $a$, the leftmost leaf $l_a$, such that $l_a$ is marked with $a$ and $l_a$ is equal to $l$ or follows $l$ (in the left-to-right order of leaves).
If \(l\) is marked with \(a\), return \(l\). Otherwise, we look for the leftmost leaf marked with \(a\) that strictly follows \(l\): We start at \(l\) and move upward in \(\mathcal T\). At each visited parent \(\beta\), let \(\gamma\) be the child from which we arrived. If \(a\notin \cE(\beta)\), we continue upward. Otherwise, using \(\cP_a(\beta)\), we find the smallest stored label strictly greater than the label of \(\gamma\). If no such label exists, we continue upward. Otherwise, let \(\gamma'\) be the corresponding child. We descend into \(\gamma'\), repeatedly choosing the child with the minimum label in \(\cP_a\) at the current node, until reaching a leaf. This leaf is the required successor. If the upward search reaches the root without finding a suitable child, no such successor exists.

The upward search selects the first subtree containing an \(a\)-marked leaf to the right of \(l\), and the downward search selects its leftmost such leaf. Both traversals take time proportional to the height of \(\mathcal T\). Since \(\mathcal T\) contains \(O(\log^5 n)\) leaves and has branching parameter \(r=\Theta(\log n)\), its height is \(O(1)\). Thus the query takes \(O(1)\) time. Predecessor queries are answered symmetrically.

% Suppose that we want to find, for a leaf $l$ and for a symbol $a$,  the leftmost leaf $l_a$, such that $l_a$ is marked with $a$ and $l_a$ follows $l$ (in the left-to-right order of leaves). Our procedure starts in the root node $\beta_R$ of $\cT$. If $\cE(\beta_R)$ does not contain $a$, then there are no leaves that are marked with $a$ in the subtree $T(v)$. Otherwise we use $P_a(\alpha)$ to find the successor $\beta_s$ of $\beta_{anc}$ where $\beta_{anc}$ is the child of $\beta_R$ that is an ancestor of $l$. If there is no successor sibling of $\beta_{anc}$ that is marked with $a$, then there is no successor of $l$ that is marked with $a$.  Otherwise we set $\beta=\beta_s$ and repeat the search:  If $\beta$ is an ancestor of $l$, we set $\beta_{anc}$ to the child of $\beta_R$ that is an ancestor of $l$; otherwise $\beta_{anc}$ is the leftmost child of $\beta$. Then we find the successor $\beta_s$ of $\beta_{anc}$. If $\beta_s$ is a leaf node, then $l_s=\beta_s$. Otherwise we set $\beta =\beta_s$.
% This procedure visits $O(1)$ nodes because at every level of $\cT$ only one node is visited. Hence it answers $O(1)$ successor queries and the total query time is $O(1)$.
% Using a symmetric procedure, we can also find the rightmost leaf that precedes $l$ and that is marked with $a$. 

When a new leaf $l$ is inserted into $T(v)$, we insert the corresponding leaf $\ell$ into $\cT$. If the new leaf is marked with a symbol $a$, we update the data structures $\cE(\alpha)$ and $\cP_a(\alpha)$ for all ancestors $\alpha$ of $\ell$. When an existing leaf $l$ receives a mark $a$, we likewise update $\cE(\alpha)$ and $\cP_a(\alpha)$ for all ancestors $\alpha$ of the corresponding leaf of $\cT$; this takes $O(1)$ time.
We maintain the property that every internal node $\alpha$ at level $i$ has at most $4r^i$ leaf descendants. We follow the convention that leaves are at level $0$ (in the weight-balanced B-tree all leaf nodes have the same depth). 
Suppose that the weight of a node $\alpha$ equals $4r^{i}$. We split $\alpha$ into two nodes using the following procedure. 
A new node $\alpha'$ is created such that $\mathtt{parent}(\alpha')=\mathtt{parent}(\alpha)$.
Children of $\alpha$ are divided into two consecutive groups, so that the difference in total weights between two groups is minimized. Nodes in the right group become the children of $\alpha'$ and nodes in the left group remain the children of $\alpha$.
Whenever a node is split, we reassign the child labels in both resulting nodes: the \(j\)-th child of $\alpha$ (resp.\ $\alpha'$) receives label \(j n^2\). We rebuild the predecessor structures $\cP_a(\alpha)$ and $\cP_a(\alpha')$ using these labels.
 Data structures $\cE(\alpha)$ and $\cE(\alpha')$ are also updated as needed. We also update the data structures for the parent $\beta$ of $\alpha$. Let $\num_\beta(\alpha)$ denote the label of a child $\alpha$ of $\beta$. We assign the label $\num_{\beta}(\alpha')=l_1+\frac{l_2-l_1}{2}=\lfloor \frac{l_1+l_2}{2} \rfloor$ where $l_1=\num_\beta(\alpha)$ and $l_2=\num_{\beta}(\alpha'')$; $\alpha$ and $\alpha''$ are children of $\beta$ that precede and follow the new child $\alpha'$. If $\alpha$ is the rightmost child, we set $l_2$ to the rightmost sentinel label . All data structures $\cP_a(\beta)$ are updated to reflect this change of labels. The total number of updates on all auxiliary data structures is bounded by $O(r^i)$. Each update of $\cE(\cdot)$ and $\cP_{\cdot}(\cdot)$ takes $O(1)$ time, so a split takes $O(r^i)$ time in total. We split a node once after a series of $\Theta(r^i)$ insertions. Hence the amortized cost of splitting a node is $O(1)$. The procedure can be de-amortized as explained in~\cite{ArgeV03}.

 The maximal node degree in the weight-balanced B-tree with parameter $r$ does not exceed $4r$. Hence the number of children that are added to a node $\alpha$ between two consecutive label resets is at most $4r=2(\log_2 n)$. When a node $\alpha$ is created, we assign its $j$-th child the label $j\cdot n^2$. Hence the difference between labels assigned to sibling nodes is $n^2$.
After \(q\) midpoint insertions since the last label reset, the minimum gap between consecutive labels, including the sentinels, is at least \(\lfloor n^2/2^q\rfloor\). Before each permitted insertion, \(q<2\log_2 n\), so this gap is at least \(2\). Thus the integer midpoint lies strictly between its two neighboring labels.

% Hence we can insert $2\log n$ nodes and assign their labels using our midpoint strategy: after $q< 2\log_2 n$ consecutive insertions between two consecutive siblings, the minimal distance between nodes exceeds $\lfloor \frac{n^2}{2^{q}}\rfloor\geq 2$. 

%we create a new node $\alpha'$ and $\alpha'$ and $\alpha''$, we insert a
%new node $\alpha''$ 
%When a new child $\alpha'$ is inserted into the tree, we assign it a label %$\alpha^p+\frac{\alpha^s-\alpha^p}{2}=\frac{\alpha^p+\alpha^s}{2}$ where $\alpha^p$ and $\alpha^s$
%are the children of $\alpha$ that precede and follow $\alpha'$. We also re-build

\paragraph{Leaf Centroid Paths.}
We also need some additional information about centroid paths that are leaves in the centroid-path tree. The dictionary $\Weiner(h)$ contains information about nodes on $h$: for a symbol $a$, we identify the lowest node $u_a$ on $h$ that has at least one leaf descendant $l_a$ marked with $a$; we set $\Weiner(h)[a]=l_a$ (any such leaf can be used). If no node on \(h\) has a leaf descendant marked with \(a\), then \(\Weiner(h)[a]=\mathtt{NULL}\).

\paragraph{Insertion of Shallow Suffixes.}
Suppose that $\wlink(h_R,A[d-i_r+1..d])=\mathtt{NULL}$ for some index $i_r$, such that $1\le i_r\le d$.
Then the lowest location in the big subtree that is labeled with a prefix of $S_{i_r}=A[d-i_r+1..d]S_{\init}$ is labeled with a prefix of $A[d-i_r+1..d]$. We can find this location in $O(d)$ time by traversing the PST. If $S_{i_r}$ continues from this location into a small subtree, we insert $S_{i_r}$ into this small subtree as described in Section~\ref{subsec:small-subtr}, with $l=l_{i_r-1}$. Otherwise, $S_{i_r}$ is inserted as a new child at this location, as in Section~\ref{subsec:appendI}. 

For $i=i_r,i_r+1,\ldots,d-1$, we insert $S_{i+1}=a_{d-i}S_i$ as follows, until all suffixes are inserted or the case described in the next paragraph occurs. Let \(T(z_i)\) be the small subtree containing \(l_i\). Using
\(D(z_i)\), we find the largest leaf \(l_p\) in \(T(z_i)\) marked
with \(a_{d-i}\) such that \(l_p\le l_i\), and the smallest leaf
\(l_s\) in \(T(z_i)\) marked with \(a_{d-i}\) such that
\(l_s\ge l_i\).
%We find the largest leaf $l_p$ such that $l_p$ is marked with $a_{d-i}$ and $l_p\le l_i$. We also find the smallest  leaf $l_s$ such that $l_s$ is marked with $a_{d-i}$ and $l_s\ge l_i$. If either $l_p$ or $l_s$ exist, we select $l_{\max}$ to be either $l_p$ or $l_s$, so that
%$\strdepth(\lca(l_{\max},l_i))=\max(\strdepth(\lca(l_p,l_i)),\strdepth(\lca(l_s,l_i)))$.
If both \(l_p\) and \(l_s\) exist, we choose \(l_{\max}\) from
\(\{l_p,l_s\}\) to maximize
\(\strdepth(\lca(l_{\max},l_i))\),
breaking ties arbitrarily. If exactly one of \(l_p\) and \(l_s\)
exists, we choose that leaf as \(l_{\max}\).
If neither $l_p$ nor $l_s$ exists, then there are no leaves in $T(z_i)$ that are marked with $a_{d-i}$. In this case we visit the lowest big ancestor $w$ of $l_i$ and consider its centroid path $h_w$. If $h_w$ is a leaf path in the centroid-path tree and $\Weiner(h_w)[a_{d-i}]\not=\mathtt{NULL}$, we set $l_{\max}=\Weiner(h_w)[a_{d-i}]$. In both cases (whether $l_{\max}$ was found in $T(z_i)$ or with $\Weiner(h_w)$), we follow the Weiner link from $l_{\max}$ to the leaf $l'_{\max}$ and find its ancestor $\nu_i$ with string depth $\strdepth(\lca(l_{\max},l_i))+1$. We create a node at $\nu_i$ if necessary and insert the leaf that holds $S_{i+1}$ as its child.  

Suppose there are no leaves in $T(z_i)$ that are marked with $a_{d-i}$ and either $\Weiner(h_w)[a_{d-i}]=\mathtt{NULL}$ or $h_w$ is not a leaf path. Let \(u\) be the lowest ancestor of \(l_i\) that has a leaf descendant
marked with \(a_{d-i}\). If no such ancestor exists, the new suffix
\(S_{i+1}\) has no nonempty common prefix with an existing suffix and
is inserted as a new child of the root. Otherwise, the path containing
\(u\) is not a leaf path in the centroid-path tree. Hence
\(\weight(u)\ge 2\tau\), and
\(\lab(u)\in\mathcal L\).

%In this case, $\lca(l_{\max},l_i)$ is some node $u\in \cL$: $u$ is the lowest ancestor of $l_i$ that has a leaf descendant marked with $a_{d-i}$. The path of $u$ is not a leaf path in the centroid tree. Hence, the weight of $u$ is at least $2\tau$.  %nodes on $h_w$ do not have leaf descendants that are marked with $a_{d-i}$. Hence the node $u$ is an ancestor of $\head(h_w)$ and its weight is at least $2\tau$.

Since $A[d-i_r+1..d]$ occurs fewer than $2\tau$ times in $T$ and is a suffix of $A[d-i+1..d]$, we have $|\lcp(S_i,\cL)|< i$.
Hence
$\strdepth(u)< i$ because the label of $u$ is in $\cL$ and $u$ is an ancestor of $l_i$. Since $|\lcp(S_{i+1},\cS)|=\strdepth(u)+1$, $|\lcp(S_{i+1},\cS)|< i+1$.
By Lemma~\ref{lemma:lcp-short}(b), $|\lcp(S_j,\cS)|< j$ for all $j\geq i+1$. In this case we insert the remaining suffixes $S_{i+1},\ldots,S_d$ in a \emph{final phase}, described in the rest of this section.

% BEGIN EDIT: Section 8, item 4 (LCPs in O(d) time)

We can find $|\lcp(S_j,\cS)|$ for all $j\ge i+1$ in $O(d)$ time as follows; here $\cS$ denotes the suffixes in the PST when the final phase begins. Since $|\lcp(S_j,\cS)|<j$, we have $|\lcp(S_j,\cS)|=|\lcp(A[d-j+1..d],\cS)|$. For $k=1,\ldots,d-i$, let $t_k=|\lcp(A[k..d],\cS)|$ and let $\mu_k$ be the location labeled with $A[k..k+t_k-1]$; we also keep a leaf $x_k$ below $\mu_k$. For $k=1$ we start at the root. For $k>1$, let $y$ be the leaf that holds the suffix starting one position after the suffix of $x_{k-1}$. The location $\nu_k$ labeled with $A[k..k+t_{k-1}-2]$ is the ancestor of $y$ at string depth $\max(t_{k-1}-1,0)$; if it lies in the big subtree, we find its centroid path with the path-ancestor data structure of Section~\ref{sec:path-ancestor} in $O(1)$ time. From $\nu_k$ we compare the following symbols of $A[k..d]$ one by one, using Lemma~\ref{lemma:pathlabel} on centroid paths and the explicit topology of small subtrees, until the first mismatch; this gives $\mu_k$ and $t_k$. As $x_k$ we take $\leaf(h)$ if $\mu_k$ lies on a centroid path $h$, and a stored leaf descendant of the next node otherwise. Step $k$ compares at most $t_k-t_{k-1}+2$ symbols, so the total time is $O(t_{d-i}+d)=O(d)$.
% END EDIT: Section 8, item 4

\paragraph{Matches Between Newly Added Suffixes.}
The searches described above find the longest common prefixes with
suffixes in a fixed reference tree. We must also account for common
prefixes between the new suffixes themselves. We retain the notation
\(S_j=A[d-j+1..d]S_{\mathrm{init}}\), where \(d\) symbols are
prepended in total. Suppose that \(S_{i+1},\ldots,S_d\) remain
to be inserted when the final phase begins. Let \(\mathcal S_0\)
be the set of suffixes represented by the reference tree at that
time; this set includes the suffixes inserted before the final phase.
For \(i+1\le j\le d\), let
\[
H_j=\lvert\lcp(S_j,\mathcal S_0)\rvert.
\]
We use the shallow-match bound \(H_j<j\le d\). Since $H_j<j$, we have $H_j=|\lcp(A[d-j+1..d],\mathcal S_0)|$; this is the value $t_{d-j+1}$ computed in the preceding paragraph.
The procedure in the preceding paragraph computes all values \(H_j\)
and the corresponding locations \(\lambda_j\) in the reference tree.
We explain how to combine this information with common prefixes
between the remaining new suffixes.

First, any two distinct remaining suffixes have a common prefix
of length less than \(2d\). Suppose, to the contrary, that two
such suffixes, starting at positions \(p<q\) in
\(AS_{\mathrm{init}}\), agree on their first \(2d\) symbols.
Put \(\delta=q-p\), and let \(t\) be the smallest positive integer
such that \(r=p+t\delta>d\). Repeatedly applying the equality
between the substrings starting at \(p\) and \(q\), we find that
the suffixes starting at \(p\) and \(r\) agree for at least
\[
2d-(t-1)\delta
\ge 2d-(d-p)
\ge d
\]
symbols. Since \(r>d\), the latter suffix is a suffix of
\(S_{\mathrm{init}}\) and therefore belongs to \(\mathcal S_0\).
This contradicts the shallow-match bound for the suffix starting
at \(p\).

We construct an auxiliary suffix tree for
\[
C=A\cdot
S_{\mathrm{init}}[1..\min(\lvert S_{\mathrm{init}}\rvert,2d)],
\]
using a fresh terminal symbol. We retain only the leaves corresponding
to \(S_{i+1},\ldots,S_d\), and suppress nonroot nodes with only one
child. Let \(\mathcal U\) denote the resulting compact trie.
The preceding bound guarantees that \(\mathcal U\) contains all
branching nodes between these full suffixes: truncation affects
only their final leaf edges. Since \(\lvert C\rvert=O(d)\), the
tree \(\mathcal U\) has \(O(d)\) nodes and can be constructed
in \(O(d)\) time. For an integer alphabet, we first assign
order-preserving ranks to the distinct symbols using dynamic fusion
trees, in \(O(d)\) total time, and then apply a linear-time
suffix-tree construction~\cite{Farach97} to the resulting alphabet of size \(O(d)\).

For each retained leaf \(j\) of \(\mathcal U\), the initial segment
of its root-to-leaf path of length \(H_j\) is already represented
in the reference tree. These segments are consistent on shared
paths. Indeed, if \(S_j\) and \(S_k\) agree for more than \(H_j\)
symbols, then \(H_k=H_j\) and \(\lambda_k=\lambda_j\): both suffixes follow
the same path in the reference tree and fail at the same next symbol.

We traverse \(\mathcal U\), storing an arbitrary descendant leaf
at every node. For a node \(x\) with representative leaf \(j\),
its label occurs in the reference tree if and only if
\(\strdepth(x)\le H_j\).
Consider an edge \((x,y)\) such that the label of \(x\) occurs
in the reference tree but the label of \(y\) does not.
For any representative leaf \(j\) below \(y\), we have
\[
\strdepth(x)\le H_j
<\strdepth(y).
\]
We cut this edge at depth \(H_j\) and attach the remaining part
of the edge, together with the subtree rooted at \(y\), at
location \(\lambda_j\). The attached subtree consists entirely of new
portions of the suffix tree. Its leaf edges are extended to
represent the full suffixes \(S_j\).

If several such subtrees attach at the same location, we
materialize that location only once. Their entering edges have
distinct first symbols; suffixes sharing the same next symbol
already belong to the same attached subtree of \(\mathcal U\).
Thus the auxiliary trie determines all additional branching
required when several new suffixes have the same insertion
location in the reference tree. Processing \(\mathcal U\) takes
\(O(d)\) time, so computing this additional topology and its
attachment locations takes \(O(d)\) total time.

\paragraph{Updates of Auxiliary Data Structures.}
We update the leaf-path dictionaries whenever a leaf \(l_i\) acquires
a mark \(a_{d-i}\) (i.e., when the leaf $l_{i+1}$ is inserted). Let \(a=a_{d-i}\), let \(w\) be the lowest big ancestor of \(l_i\), and let
\(h\) be its centroid path. If \(h\) is a leaf path, we update
\(\Weiner(h)[a]\) as follows. Together with each
stored witness, we keep the corresponding node \(u_a\). If the
entry is absent or
\(\strdepth(w)>\strdepth(u_a)\),
we set \(u_a=w\) and \(\Weiner(h)[a]=l_i\);
otherwise the entry remains unchanged. No other leaf-path
dictionary is affected, since no other leaf path contains an
ancestor of \(l_i\). Thus, provided that the centroid path
decomposition is unchanged and \(w\) is available, maintaining
these dictionaries takes \(O(1)\) time per newly marked leaf.

\commented{
Inserting into small subtrees.

Updating w-links

Updating structures for small subtrees
}

\section{Data Structure for the Buffer Text}
\label{sec:buffertext}
% BEGIN EDIT: Section 9 (replace suffix labels and P_a by D_B)
% OLD: The data structure for the text $T_B$ is the suffix tree. Every suffix $S=T[i..n_B]$ is assigned an integer label $\lab(S)$, such that $1\le \lab(S)\le n^2d$ and $\lab(S_1)< \lab(S_2)$ iff $S_1< S_2$. $T_B$ contains at most $2d$ symbols $n_B\leq 2d$. 
% OLD: 
% OLD: For every symbol $a$ that occurs in $T[1..n_B]$, we keep a data structure $P_a$. $P_a$ contains the labels of all suffixes $T[i..n_B]$ such that $T[i-1]=a$; $P_a$ supports insertions and predecessor queries in $O(1)$ time~\cite{fusionTree}. We also keep Weiner links for all leaves of the suffix tree and  a data structure that answers lowest common ancestor queries on the suffix tree. 
% OLD: 
% OLD: 
% OLD: When a new symbol $a$ is pre-pended to the text, a new suffix $T[1..n_B]=aT[2..n_B]$ must be inserted into the suffix tree. We find the predecessor and the successor of $\lab(T[2..n_B])$ in $P_a$. 
% OLD: Let $l_p$, $l_m$ and $l_s$ denote the leaves holding the suffixes that correspond to  the predecessor of $\lab(T[2..n_B])$, $\lab(T[2..n_B])$,  and the successor of $\lab(T[2..n_B])$ respectively. We find $t_1=\strdepth(\lca(l_p,l_m))$ and $t_2=\strdepth(\lca(l_s,l_m))$.
% OLD: In the special case when $a$ did  not occur in $T[2..n_B]$, we initialize the new structure $P_a$ and add $T[1..n_B]$ as a new child of the root node.
% NEW:
The data structure for the buffer text $T_B$ is its suffix tree.
 We store all its suffix-tree leaves, in lexicographic order, in one instance $D_{\mathrm{buf}}$ of the data structure {$D(v)$} described in Section~\ref{subsec:main-shallow}, paragraph ``Data Structure for Small Subtrees.'' The leaf representing $T_B[i..n_B]$ is marked with $T_B[i-1]$ for $i>1$; the leaf representing the entire buffer is unmarked. For simplicity, we assume that the text ends with the terminal \(\$\) symbol, so that \(\$>a\) for any other symbol $a$ in $T_B$. The buffer text contains at most $2d$ ordinary symbols, so that $n_B\le 2d+1$. Since there are {$O(d)=O(\log n)$} leaves, predecessor and successor queries restricted to a given mark, as well as insertions, take {$O(1)$} time. Data structure $D_{\mathrm{buf}}$ works on a list $L_{\mathrm{buf}}$ that contains all suffixes of $T_B$ sorted in lexicographic order.
We also keep Weiner links for all leaves of the suffix tree and a data structure that answers lowest common ancestor queries on the suffix tree (Section~\ref{sec:path-ancestor}).

When a new symbol {$a$} is prepended, let {$n_B$} denote the buffer length after this insertion. We must insert the new suffix $T_B[1..n_B]=aT_B[2..n_B]$. Let $l_m$ be the existing leaf representing $T_B[2..n_B]$. Before marking {$l_m$} with {$a$}, we query $D_{\mathrm{buf}}$ for its predecessor {$l_p$} and successor {$l_s$} marked with {$a$}. We set {$t_1=\strdepth(\lca(l_p,l_m))$} and {$t_2=\strdepth(\lca(l_s,l_m))$} whenever the corresponding leaf exists; for a missing predecessor or successor, we set the corresponding value to {$-1$}.

  If neither $l_p$ nor $l_s$ exists, this means that \(a\) did not occur in \(T_B\) before this insertion and no existing leaf is marked with \(a\). We create a new leaf \(l_{\mathtt{new}}\), representing $aT_B[2..n_B]$, as a child of the root and insert it into $L_{\mathrm{buf}}$ in lexicographic order, using the method that will be explained later. We then mark \(l_m\) with \(a\) and create the Weiner link labeled with \(a\) from \(l_m\) to \(l_{\mathtt{new}}\). The new leaf $l_{\mathtt{new}}$ remains unmarked. The data structure $D_{\mathrm{buf}}$ is updated accordingly.

% END EDIT: Section 9 (replace suffix labels and P_a by D_B)
% OLD: Let $l'_p$ and $l'_s$ be the leaves that are reached by following Weiner links from  $l_p$ and $l_s$ respectively.
If at least one of $l_p$ and $l_s$ exists, we proceed as follows.
For each existing leaf {$l_p$} or {$l_s$}, let {$l'_p$} or {$l'_s$}, respectively, be the target of its Weiner link labeled with {$a$}.
%We denote by $u$ the lowest common ancestor of $l'_p$ and $l'_s$. %Let $u_p$ and $u_s$ be the children of $u$ that are ancestors of $l'_p$ and $l'_s$ respectively.
Suppose that $t_1>t_2$. We find the lowest ancestor $v$ of $l'_p$ such that $\strdepth(v)\leq t_1+1$. If $\strdepth(v)=t_1+1$, we set $v'=v$. Otherwise let \(v_p\) be the child of \(v\) that is an ancestor of \(l'_p\). We split the edge \((v,v_p)\) at string depth \(t_1+1\) and create a new node \(v'\). Then we create a new leaf $l_{\mathtt{new}}$ as a child of $v'$. The case when $t_1\le t_2$ is processed in the same way.
% OLD: Finally we create a new Weiner link from the leaf that stores $T[2..n_B]$ to $l_{\mathtt{new}}$.
% BEGIN EDIT: Section 9 (common D_B and Weiner-link updates)
In either of these two cases, we create the Weiner link labeled with {$a$} from {$l_m$} to {$l_{\mathtt{new}}$}, insert $l_{\mathtt{new}}$ into $L_{\mathrm{buf}}$ at its lexicographic position, immediately after $l'_p$ (or immediately before $l'_s$), and mark {$l_m$} with {$a$}. The new leaf is initially unmarked. We also update the data structure $D_{\mathrm{buf}}$ accordingly. To add the mark to {$l_m$}, we update the dictionaries and predecessor structures at its {$O(1)$} ancestors in the balanced tree underlying $D_{\mathrm{buf}}$, as in the insertion procedure for {$D(v)$}. Thus these updates take {$O(1)$} time.
% END EDIT: Section 9 (common D_B and Weiner-link updates) 

It remains to describe a data structure that finds the node $v$. We visit all nodes of the suffix tree in the order induced by Euler tour: we visit the root node $u_R$, then we recursively visit the subtree rooted at the leftmost child of $u_R$, then we visit the root node again, then we recursively visit the subtree rooted at the second child from the left, and so on. When the rightmost subtree of $u_R$ is visited, we visit $u_R$ for the last time and complete the tour. The nodes are stored in this order in a list $L_E$. The same vertex can appear multiple times in $L_E$. A node $u$ occurs in $L_E$ once more than the number of its children. Hence the size of $L_E$ is $O(d)$ and we insert $O(1)$ elements into $L_E$ when a new suffix is prepended. Every element of $L_E$ is assigned a label. Every occurrence of any suffix tree node $u$ in $L_E$ is represented by a two-dimensional point $p$, so that the $x$-coordinate of $p$ is the label of that occurrence in $L_E$ and its $y$-coordinate is the string depth of $u$.
The lowest ancestor $v$ of a leaf $l$ with string depth at most $t$ corresponds to the point $p(v)$ with the smallest $x$-coordinate among all points $p$ in $[l_x,+\infty) \times (-\infty, t]$, where $l_x$ is the label of $l$.
To see this, suppose that \(\operatorname{strdepth}(l)>t\), and let \(v_p\) be the child of \(v\) on the path to \(l\). Every occurrence between \(l\) and the next occurrence of \(v\) belongs to the subtree rooted at \(v_p\). By the choice of \(v\), all nodes in this subtree have string depth greater than \(t\). Consequently, the first qualifying occurrence represents \(v\). Its immediately preceding element represents \(v_p\), because the Euler tour returns from \(v_p\) to \(v\) after visiting all nodes in the subtree of $v_p$.
There is a data structure that answers such queries and supports insertions of points in $O(1)$ time; this data structure is described in Section~\ref{sec:path-ancestor}, paragraph ``Geometric Data Structure''.
When a symbol is prepended to $T_B$, we add one leaf and at most one internal node to the suffix tree. This incurs $O(1)$ insertions into the list $L_E$. 

The labeling of elements in $L_E$ uses the same method that was employed to label children of internal nodes in the data structure $D(v)$ in
Section~\ref{par:datasmall}. When $L_E$ is created for a new instance, which contains only the symbol \$, we set $\num(-\infty)=0$ and $\num(+\infty)=(2n)^4$ for two sentinel elements in $L_E$.  
When a new element $e$ is inserted, we assign it the label $\num(e)=\lfloor\frac{\num(p_e)+\num(s_e)}{2}\rfloor$, where $p_e$ is the predecessor of $e$ in $L_E$ and $s_e$ is the successor of $e$ in $L_E$.
Each symbol insertion adds at most four elements to \(L_E\). Hence an instance undergoes at most \(8d+4\) label insertions, including initialization. Since \(d=\frac12\log_2 n\), the initial gap \(16n^4\) is sufficient for all midpoint insertions, and every label occupies \(O(\log n)\) bits.

Now we explain how the lexicographic list of suffixes is maintained when a symbol previously absent from $T_B$ is prepended. We keep a predecessor structure on the symbols that occur in \(T_B\). For each such symbol \(b\), we store pointers to the first and last leaves in $L_{\mathrm{buf}}$ whose suffixes begin with \(b\). When a previously absent symbol \(a\) is prepended, we find the largest stored symbol \(b<a\) and insert \(l_{\mathtt{new}}\) immediately after the last leaf whose suffix begins with \(b\). If no such symbol exists, we insert \(l_{\mathtt{new}}\) at the beginning of $L_{\mathrm{buf}}$. We then insert \(a\) into the predecessor structure and set both its pointers to \(l_{\mathtt{new}}\). When a symbol \(a'\) that already occurred in $T_B$ is prepended, we update its first or last pointer if the new leaf extends the corresponding boundary. Since the structure contains \(O(d)=O(\log n)\) symbols, we implement it using a fusion tree~\cite{FredmanW93,PatrascuT14} that supports queries and updates in \(O(1)\) time.

Symbols from the buffer text are transferred to the main text $T_M$ in blocks of $d$ symbols.
At the beginning of each transferring round, we create a new instance $T_B^N$ of $T_B$ and initialize $T_B^N$ to the length-$1$ string that consists of a single terminal symbol \(\$\) (representing an empty sequence of ordinary symbols). When a new symbol is added to the text, we prepend it to both $T_B$ and $T_B^N$. When all $d$ symbols have been transferred to $T_M$, we set $T_B:=T_B^N$ and discard the old instance of $T_B$. Each round lasts until $d$ new symbols have arrived; hence, at the beginning of a round, the ordinary symbols of $T_B$ are exactly the $d$ symbols that are transferred in this round.

\section{Marginal Occurrences}
\label{sec:marginal}

%Occurrences that overlap with both the main text and the buffer text. Weiner links on the buffer text and suffix links on the main text.

An occurrence of $P$ is \emph{marginal} if it starts in $T_B$ and ends in $T_M$. Let $n_b=n_B-1$ denote the number of ordinary symbols in $T_B$.
For every $i$, $1\le i \le \min(n_b,|P|-1)$, we check if $P[1..i]$ is a suffix of the
buffer text and $P[i+1.. |P|]$ is a prefix of the main text. This can be done as follows. 

We identify the longest prefix $P[1.. k]$ of $P$ that occurs in the buffer text and its location $v_k$ in the suffix tree of $T_B$. We traverse the path from the root to $v_k$ and find the locations $v_i$ where $v_i$ is the location of $P[1..i]$, $1\le i\le k$. Then for every $i$, $1\le i\le k$, we check if the leaf that holds the suffix $T_B[n_b-i+1..n_B]$ of the buffer text is a descendant of $v_i$. All such indexes $i$ are added to the list $L_1$ in increasing order.

Next we turn to the main text. Let $f$ denote the number of symbols
from $T_B$ that have already been added to $T_M$. For $i\ge 1$, let
$u_i$ denote the location of the longest prefix of $P[i..|P|]$ that
occurs in the main text. We find $u_1$ as in
Section~\ref{sec:queries-main}. To obtain $u_{i+1}$ from $u_i$, we
choose a leaf $x$ below $u_i$ and find the ancestor of $\slink(x)$ at
string depth $\max(|\lab(u_i)|-1,0)$; if it lies in the big subtree,
we find its centroid path with the path-ancestor data structure of
Section~\ref{sec:path-ancestor} in $O(1)$ time. From this location,
we compare the following symbols of $P$ one by one, using
Lemma~\ref{lemma:pathlabel} on centroid paths and the explicit
topology of small subtrees, until the first mismatch; this gives
$u_{i+1}$. As in the computation of matching statistics, the total
number of comparisons is $O(|P|)$, so all locations $u_i$ are found
in $O(|P|)$ time. Let $s$ be the smallest index such that
$P[s..|P|]$ occurs in the main text; then $u_i$ is the location of
$P[i..|P|]$ for every $i\ge s$. Let $l_{f+1}$ be the leaf that holds
$T[f+1..n]$. We add to the list $L_2$, in increasing order, all
indices $i$, $s\le i\le\min(k+1,|P|)$, such that $u_i$ is an
ancestor of $l_{f+1}$. Thus $L_2$ contains all these indices $i$ such
that $P[i..|P|]$ is a prefix of $T[f+1..n]$.

If an index $i$ occurs in $L_1$ and $i+1$ occurs in $L_2$,
$P[1..i]$ is a suffix of the buffer text and $P[i+1..|P|]$ is a prefix of the main text (ignoring the first $f$ symbols that were moved from the buffer text to the main text by the background process). 
Hence there is an occurrence of $P$ that starts at position $n_b-i+1$ of the buffer text.
We can construct $L_1$ in time $O(k)=O(|P|)$ and $L_2$ in time $O(|P|+k)=O(|P|)$. We can examine $L_1$ and $L_2$ and report all necessary indexes $i$ in time $O(k)=O(|P|)$. Thus all marginal occurrences of $P$ can be reported in time $O(|P|)$.

% BEGIN EDIT: Path-ancestor data structures
\section{Auxiliary Data Structures}
\label{sec:path-ancestor}
\paragraph{Path Dictionaries.}
The dictionary $\Dict_h$ supports membership queries, insertions, deletions, and listing all nodes on the centroid path $h$.
Its implementation consists of two components. The dictionary structure supports insertions and deletions in $O(1)$ time w.h.p.\ and membership queries in $O(1)$ worst-case time~\cite{DemaineMPP06}. The second component is a doubly-linked list that stores pointers to all nodes on $h$. Dictionary entries contain pointers to list elements, so that a node pointer can be deleted from a list in $O(1)$ time. All other dictionaries used in this paper, such as $\child_u$, $\Weiner(h)$, and the w-dictionaries, are implemented in the same way. Since each new letter causes $O(1)$ dictionary operations, a union bound shows that, for any constant $c>0$, a new letter is processed in $O(1)$ time with probability at least $1-n^{-c}$. Since every word used by our data structures, including the dictionaries, is written by some operation, and each new letter is processed in $O(1)$ time, the data structure uses $O(n)$ words of space.

\paragraph{Path-Ancestor Structures.}
For every centroid path $h$, we store at $\head(h)$ a fusion tree containing the keys $\strdepth(\parent(\head(g)))+1$ for all ancestor paths $g$ of $h$, including $h$ itself; the key of $g$ is the smallest string depth of a location on $g$, since the edge from $\parent(\head(g))$ to $\head(g)$ belongs to $g$ (the key of the root path is $0$). Each entry also stores a pointer to {$g$}. Since a centroid path has {$O(\log n)$} ancestor paths, predecessor queries, insertions, and deletions take {$O(1)$} time. Given a depth $t$, a predecessor query identifies the ancestor path with the greatest key not exceeding $t$, i.e., the path that contains the location at string depth $t$ on the path from the root to $\head(h)$, or reports that no such path exists.

\paragraph{W-Dictionaries.}
In every centroid path $h$, we also store the labels and targets of its w-links in a w-dictionary. The w-dictionary of $h$ is a trie on the reverses of the strings $s$; a node of the trie that corresponds to the reverse of $s$ is marked if there is a w-link from $h$ labeled with $s$, i.e., $\wlink(h,s)\neq\mathtt{NULL}$, and a marked node stores the target $\wlink(h,s)$. Since $s_i=a_{d-i+1}s_{i-1}$, the reverse of $s_i$ extends the reverse of $s_{i-1}$ by one symbol; hence the trie node of $s_i$ is a child of the trie node of $s_{i-1}$, and it is found in $O(1)$ time using a dictionary on the children of each trie node. 

\paragraph{Lowest Common Ancestors.}
We maintain a data structure that answers lowest common ancestor
queries on the centroid-path tree in $O(1)$ time. For a node $x$
of the PST, let $w_x=x$ if $x$ is big, and let $w_x$ be the lowest
big ancestor of $x$ otherwise; let $h_x$ be the centroid path
that contains $w_x$.
Consider two nodes $x$ and $y$ that do not belong to the same
small subtree, and let $g$ be the lowest common ancestor of
$h_x$ and $h_y$ in the centroid-path tree.
For $z\in\{x,y\}$, let $e(z)$ be the string depth at which the
path from the root to $z$ leaves $g$: if $h_z=g$, then
$e(z)=\strdepth(w_z)$; otherwise, a successor query for the key of
$g$ in the path-ancestor structure of $h_z$ returns the child
$g_z$ of $g$ that is an ancestor of $h_z$, and
$e(z)=\strdepth(\parent(\head(g_z)))$.
Then $\lca(x,y)$ is the node of $g$ with string depth
$\min(e(x),e(y))$, which we find with $\Dict_g$.
If $x$ and $y$ belong to the same small subtree, we use a
dynamic LCA structure~\cite{ColeH05} on that subtree, whose
topology is stored explicitly.
In both cases, a query takes $O(1)$ time.

\paragraph{Geometric Data Structure.}
In this paragraph, we explain how the geometric data structure needed to support weighted level ancestor queries in Section~\ref{sec:buffertext} can be implemented.
The data structure stores a set $V$ of two-dimensional points so that, for a query range $Q=[q_x,+\infty)\times(-\infty,q_y]$, we can find the leftmost point (i.e., the point with the minimal $x$-coordinate) in $Q\cap V$. 
We will call such queries two-sided range successor queries.

\begin{lemma}
  \label{lemma:2s-tbl}
  Let $m=\log^{1/2}n$. There exists a data structure that answers two-sided range successor queries on a set of at most $m$ points with integer coordinates. Queries are answered in $O(1)$ time. The data structure uses a universal look-up table of size $o(n)$ that can be constructed in $o(n)$ time. 
\end{lemma}
\begin{proof}
  We store points of $V$ in the rank space. In the rank space, there are $m^2$ different points. 
    Since a set contains at most $m$ points, there are at most $m^{2m}$ different sets. There are $m^2$ different queries that can be asked. 
  We keep a look-up table that stores answers to all possible queries on all possible data sets: $\mathrm{Tbl}_Q[V][i][j]$ stores the answer to a query $[i,+\infty)\times (-\infty,j]$ on a set $V$. Hence $\mathrm{Tbl}_Q$ uses $O(m^{2m}\cdot m^2)=O(m^{2m+2})=o(n)$ space.

The data structure for a set $V$ is the index of $V$, which takes $O(m\log m)=o(\log n)$ bits and fits into one word. Since $n$ grows, we rebuild the universal table in the background when $n$ doubles; the value of $m$ is fixed between rebuilds. 
\end{proof}

The data structure from Lemma~\ref{lemma:2s-tbl} can be dynamized.
\begin{lemma}
  \label{lemma:2s-tbl-dyn}
  Let $m=\log^{1/2}n$. There exists a data structure that answers two-sided range successor queries on a set of at most $m$ points with integer coordinates. Queries are answered in $O(1)$ time and updates are supported in $O(1)$ time. The data structure uses a universal look-up table of size $o(n)$ that can be constructed in $o(n)$ time. 
\end{lemma}
\begin{proof}
  We assign every point in $V$ an integer id, $\mathrm{id}(p)$. $\mathrm{id}(p)$ is an integer from $[1,m]$. We keep the maximum id value $\mathrm{maxid}$ for any point stored in the data structure. We also keep a list of all vacant ids, i.e., the list of all integers $a$ in $[1,\mathrm{maxid}]$ such that $a\not=\mathrm{id}(p)$ for any $p\in V$. We also store a dictionary of all used ids that maps ids to points.
  
  We store $x$- and $y$-coordinates of points in the rank space and we treat $\mathrm{id}(p)$ as an auxiliary point coordinate. Hence there are $m^3$ different points. The $x$- and $y$-coordinates of points in $V$ are kept in two fusion trees~\cite{FredmanW93,PatrascuT14}, so that a coordinate is reduced to the rank space, and a new coordinate is inserted, in $O(1)$ time. Since $V$ contains at most $m$
  points the total number of different sets is $O(m^{3m})$. We can ask $m^2$ different queries. Hence, the look-up table $\mathrm{Tbl}_Q$ that stores the answers to all possible queries on all possible sets, uses $O(m^{3m}\cdot m^2)=o(n)$ space.  The answer stored in $\mathrm{Tbl}_Q[V][i][j]$ is the id of the reported point; using the dictionary of ids, we can obtain the original point coordinates.

We also store a table for all possible data sets and for all possible insertions/deletions:  $\mathrm{Tbl}_I[V][p]$ stores the index of the set $V'$ obtained by inserting $p$ into $V$ and updating the ranks of all other points of $V$;  analogously $\mathrm{Tbl}_D[V][p]$ stores the index of the set $V'$, such that $V'$ is obtained by deleting the point $p$ from the set $V$. Points of $V$ are identified by their ids, each of which takes $O(\log m)=O(\log\log n)$ bits. We cannot identify points by their ranks: when a point is inserted into $V$ or deleted from $V$, the ranks of other points shift. Since all points of $V$ are identified by their ids, the ranks of all points can be updated when a point is inserted or deleted. Since there are at most $m^3$ potential points that can be inserted or deleted into a set $V$, the size of
$\mathrm{Tbl}_I$ and $\mathrm{Tbl}_D$ is $O(m^{3m}\cdot m^3)=o(n)$. 

In order to answer a query $[q_x,+\infty)\times (-\infty,q_y]$, we reduce $q_x$ and $q_y$ to the rank space and answer the query using $\mathrm{Tbl}_Q$. 
When a new point $p$ is inserted, we reduce its $x$- and $y$-coordinates to the rank space. Then we assign $\mathrm{id}(p)$ to $p$ by selecting the first element from the list of vacant ids; if the list of vacant ids is empty, we set $\mathrm{id}(p)=\mathrm{maxid}+1$ and increment $\mathrm{maxid}$ by $1$. Finally we update the set $V$ using $\mathrm{Tbl}_I$.
When a point $p$ is removed, we update $V$ using $\mathrm{Tbl}_D$ and add $\mathrm{id}(p)$ to the list of vacant ids. 
\end{proof}

\begin{lemma}
  \label{lemma:2s-polylog}
  Let $m=O(\tau)$. There exists an $O(m)$-space data structure that answers two-sided range successor queries on a set of at most $m$ points with integer coordinates. Queries are answered in $O(1)$ time and insertions are supported in $O(1)$ time.
\end{lemma}
\begin{proof}
Our data structure combines the grid approach with the data structure from Lemma~\ref{lemma:2s-tbl-dyn}. The grid is divided into $O(r)$ rows $R_i=(-\infty,+\infty)\times [y_{i-1}+1,y_i]$ and $O(r)$ columns $C_j=[x_{j-1}+1,x_j]\times(-\infty,+\infty)$ where the boundaries $x_j$ and $y_i$ are chosen so that every row and every column contains at most $2r'$ points; when a row or a column contains $\frac32 r'$ points, we start splitting it into two, and the split is completed before the row (column) contains $2r'$ points. Since every split produces rows (columns) with at least $\frac34 r'$ points, there are $O(r)$ rows and $O(r)$ columns. Here $r'=\frac{m}{r}$ and $r=\log^{1/4}n$. Let $Z_{ij}=R_i\cap C_j$ and let $\min_{ij}$ denote the point with the smallest $x$-coordinate in $Z_{ij}$. The set $V^t$ stores the points $\min_{ij}$ for every cell $Z_{ij}$. 
We store $V^t$ in the data structure of Lemma~\ref{lemma:2s-tbl-dyn}. If $r'>\log^{1/4}n$, we also store a recursively defined data structure for every $R_i$ and for every $C_j$. Otherwise we store the data structure of Lemma~\ref{lemma:2s-tbl-dyn} for each $R_i$ and each $C_j$. The recursion depth of our data structure is $O(1)$. Hence every point is stored in $O(1)$ data structures. 

Given a query $Q=[q_x,+\infty)\times(-\infty,q_y]$, we identify the column $C_j$ and the row $R_i$ that satisfy $x_{j-1}<q_x\le x_j$ and $y_{i-1}<q_y\le y_i$. The boundaries $x_j$ and $y_i$ are kept in two fusion trees~\cite{FredmanW93,PatrascuT14}; hence $C_j$ and $R_i$ are found in $O(1)$ time. We find the leftmost point $p_1$ in $(x_j,+\infty)\times(-\infty,y_{i-1}]$ using $V^t$; we also find the leftmost point $p_2$ in $R_i\cap Q$ and the leftmost point $p_3$ in $C_j\cap Q$.
Points $p_2$ and $p_3$ can be found using data structures for the row $R_i$ and the column $C_j$ respectively. The leftmost point in $Q\cap V$ is the point with the smallest $x$-coordinate among $p_1$, $p_2$, and $p_3$. Answering a query requires answering $O(1)$ queries on data structures implemented using Lemma~\ref{lemma:2s-tbl-dyn}. Hence the overall query time is $O(1)$. 

When a new point $p$ is inserted, we find the cell $Z_{ij}$ that contains $p$ and update $\min_{ij}$ and $V^t$ if necessary. Then we recursively update the data structures for $R_i$ and $C_j$.
When a row or a column is split, the data structures for the two resulting rows (resp.\ columns) are rebuilt in the background; hence a split takes $O(1)$ time per insertion. Overall an insertion requires $O(1)$ updates of data structures implemented using Lemma~\ref{lemma:2s-tbl-dyn}. Hence the overall update cost is $O(1)$. 
\end{proof}

% END EDIT: Path-ancestor data structures

\section{W-Links and New Nodes}
\label{sec:wlink-nodes}
\begin{lemma}
  \label{lemma:leaves-wlinks}
  We call a big node an \emph{lb-node} if none of its children is big. A w-link needs to be added or changed only when a new lb-node is added to the big subtree or when a new centroid path is created. 
\end{lemma}
\begin{proof}
  Suppose that a big node $u$ that is not an lb-node is added, and let $u'$ be a lowest big proper descendant of $u$. Then $\slabel(u)=s\cdot \alpha$ and $\slabel(u')=s\cdot \alpha'$ where $\alpha$ is a prefix of $\alpha'$ and $s$ is a length-$i$ prefix of $\slabel(u)$ for some $i$, $1\le i \le \min(d,\strdepth(u)-1)$. Let $v$ and $v'$ be nodes with labels $\alpha$ and $\alpha'$ respectively. Let $h=\spath(v)$ and $h'=\spath(v')$. Since $v$ is an ancestor of $v'$, the path $h$ must be an ancestor of $h'$, or equal to $h'$, in the centroid-path tree. By definition of w-links, there existed a w-link from $h'$ before $u$ was added. Hence there also existed a w-link from $h$ before $u$ was added: the location with label $s\cdot\alpha$ is an ancestor of $u'$, so it already existed on an edge of a centroid path, since all ancestors of a big node are big, and $v\in h$ has a Weiner link labeled with $s$ to this location. The target of the w-link from $h$ was either the path of $u$ or its descendant, so creating $u$ does not change $\wlink(h,s)$.

\end{proof}

A new node can also be the source of a Weiner link; for the nodes created when a suffix is inserted, this case is handled in Lemma~\ref{lemma:insert-no-wlink}.

\begin{lemma}
  \label{lemma:insert-no-wlink}
  The insertion of a suffix $S_i$ described in Section~\ref{subsec:appendI}, before any rebuilding work is performed, does not change $\wlink(g,s)$ for any centroid path $g$ and any string $s$.
\end{lemma}
\begin{proof}
  The insertion creates the leaf $l_i$ and possibly an internal node $u_b$. The leaf $l_i$ is small and does not belong to any centroid path. The set of big nodes and the set of centroid paths do not change, except that $u_b$ may be added to an existing centroid path. All new locations lie on the edge that leads to $l_i$; these locations do not belong to any centroid path. Hence the insertion creates no new target of a Weiner link on a centroid path, and the only possible new source is $u_b$.

  If $u_b$ is small, it does not belong to any centroid path and nothing changes. Suppose that $u_b$ is big. Then $u_b$ is created on an edge $(u,v)$, where $v$ is big and $u_b$ is added to the centroid path $h$ of $v$; thus $v$ is the child of $u_b$ on $h$. Before the insertion, the location of $\lab(u_b)$ was an implicit location on the edge $(u,v)$; hence every suffix in the PST that started with $\lab(u_b)$ also started with $\lab(v)$.

  Consider a string $s$ such that there is a Weiner link labeled with $s$ from $u_b$ to a location $x$, i.e., $\lab(x)=s\cdot\lab(u_b)$. Let $X$ be any suffix in the PST, after the insertion, that starts with $s\cdot\lab(u_b)$, and let $Y$ be the suffix obtained by removing the first $|s|$ symbols from $X$. Then $Y$ is in the PST, $Y\neq S_i$ because $Y$ is shorter than $X$, and $Y$ starts with $\lab(u_b)$. Hence $Y$ was in the PST before the insertion, so $Y$ starts with $\lab(v)$ and $X$ starts with $s\cdot\lab(v)$. Thus the location $y$ with $\lab(y)=s\cdot\lab(v)$ exists, and every suffix that passes through $x$ also passes through $y$. Therefore no node lies strictly between $x$ and $y$, so $x$ and $y$ lie on the same edge and hence on the same centroid path. Since $v\in h$, the Weiner link labeled with $s$ from $v$ already leads to this centroid path, and $\wlink(h,s)$ does not change.
\end{proof}

%\subsection*{Acknowledgements}
\paragraph{Use of Generative AI.}
% ChatGPT 6 Astra Medium and Claude Opus 5.5 were used to assist with  writing the paper and improving the presentation.  The first draft of the paper and all technical content were generated by the authors. The authors take full responsibility for the correctness and content of the paper.
ChatGPT 6 Astra Medium and Claude Opus 5.5 were used to assist with editing the paper and improving the presentation.  The authors obtained all technical results and drafted the manuscript. The authors take full responsibility for the content of the paper.

\bibliographystyle{plain}
\bibliography{realtimeindex}

@Article{ArgeV03,
  author       = {Arge, Lars and Vitter, Jeffrey Scott},
  title	       = {Optimal External Memory Interval Management},
  journal      = {SIAM Journal on Computing},
  year	       = 2003,
  volume       = 32,
  number       = 6,
  month	       = Jan,
  pages	       = {1488-1508},
  issn	       = {1095-7111},
  doi	       = {10.1137/s009753970240481x},
  url	       = {http://dx.doi.org/10.1137/s009753970240481x},
  publisher    = {Society for Industrial & Applied Mathematics (SIAM)}
}

@Article{FredmanW93,
  author       = {Fredman, Michael L. and Willard, Dan E.},
  title	       = {Surpassing the information theoretic bound with
                  fusion trees},
  journal      = {Journal of Computer and System Sciences},
  year	       = 1993,
  volume       = 47,
  number       = 3,
  month	       = Dec,
  pages	       = {424-436},
  issn	       = {0022-0000},
  doi	       = {10.1016/0022-0000(93)90040-4},
  url	       = {http://dx.doi.org/10.1016/0022-0000(93)90040-4},
  publisher    = {Elsevier BV}
}

@InProceedings{PatrascuT14,
  author       = {Patrascu, Mihai and Thorup, Mikkel},
  title	       = {Dynamic Integer Sets with Optimal Rank, Select, and
                  Predecessor Search},
  year	       = 2014,
  booktitle    = {2014 IEEE 55th Annual Symposium on Foundations of
                  Computer Science},
  publisher    = {IEEE},
  month	       = Oct,
  pages	       = {166--175},
  doi	       = {10.1109/focs.2014.26},
  url	       = {http://dx.doi.org/10.1109/focs.2014.26}
}

@inproceedings{Farach97,
  author    = {Martin Farach},
  title     = {Optimal Suffix Tree Construction with Large Alphabets},
  booktitle = {38th Annual Symposium on Foundations of Computer Science (FOCS)},
  pages     = {137--143},
  publisher = {IEEE},
  year      = {1997}
}

@inproceedings{FischerG15,
  author    = {Johannes Fischer and Pawe{\l} Gawrychowski},
  title     = {Alphabet-Dependent String Searching with Wexponential
               Search Trees},
  booktitle = {Proc. 26th Annual Symposium on Combinatorial Pattern
               Matching (CPM)},
  series    = {Lecture Notes in Computer Science},
  volume    = {9133},
  publisher = {Springer},
  year      = {2015},
  doi       = {10.1007/978-3-319-19929-0_14}
}

@article{AnderssonT07,
  author  = {Arne Andersson and Mikkel Thorup},
  title   = {Dynamic Ordered Sets with Exponential Search Trees},
  journal = {Journal of the ACM},
  volume  = {54},
  number  = {3},
  pages   = {13},
  year    = {2007},
  doi     = {10.1145/1236457.1236460}
}

@inproceedings{AmirKLL05,
  author    = {Amihood Amir and Tsvi Kopelowitz and Moshe Lewenstein and
               Noa Lewenstein},
  title     = {Towards Real-Time Suffix Tree Construction},
  booktitle = {Proc. 12th International Symposium on String Processing and
               Information Retrieval (SPIRE)},
  series    = {Lecture Notes in Computer Science},
  volume    = {3772},
  pages     = {67--78},
  publisher = {Springer},
  year      = {2005},
  doi       = {10.1007/11575832_9}
}

@inproceedings{AmirN08,
  author    = {Amihood Amir and Igor Nor},
  title     = {Real-Time Indexing over Fixed Finite Alphabets},
  booktitle = {Proc. 19th Annual ACM-SIAM Symposium on Discrete Algorithms
               (SODA)},
  pages     = {1086--1095},
  publisher = {SIAM},
  year      = {2008},
  doi       = {10.1145/1347082.1347201}
}

@inproceedings{Slisenko78,
  author    = {A. O. Slisenko},
  title     = {String-Matching in Real Time: Some Properties of the Data
               Structure},
  booktitle = {Proc. 7th Symposium on Mathematical Foundations of Computer
               Science (MFCS)},
  series    = {Lecture Notes in Computer Science},
  volume    = {64},
  pages     = {493--496},
  publisher = {Springer},
  year      = {1978}
}

@inproceedings{Kosaraju94,
  author    = {S. Rao Kosaraju},
  title     = {Real-Time Pattern Matching and Quasi-Real-Time Construction
               of Suffix Trees (Preliminary Version)},
  booktitle = {Proc. 26th Annual ACM Symposium on Theory of Computing (STOC)},
  pages     = {310--316},
  publisher = {ACM},
  year      = {1994},
  doi       = {10.1145/195058.195170}
}

@article{KucherovN17,
  author  = {Gregory Kucherov and Yakov Nekrich},
  title   = {Full-Fledged Real-Time Indexing for Constant Size Alphabets},
  journal = {Algorithmica},
  volume  = {79},
  number  = {2},
  pages   = {387--400},
  year    = {2017},
  doi     = {10.1007/s00453-016-0199-7}
}

@article{BreslauerI13,
  author  = {Dany Breslauer and Giuseppe F. Italiano},
  title   = {Near Real-Time Suffix Tree Construction via the Fringe Marked
             Ancestor Problem},
  journal = {Journal of Discrete Algorithms},
  volume  = {18},
  pages   = {32--48},
  year    = {2013}
}

@inproceedings{KopplK26,
  author    = {Dominik K{\"o}ppl and Gregory Kucherov},
  title     = {Near-Real-Time Solutions for Online String Problems},
  booktitle = {Proc. 37th Annual Symposium on Combinatorial Pattern
               Matching (CPM)},
  series    = {LIPIcs},
  volume    = {369},
  pages     = {2:1--2:17},
  publisher = {Schloss Dagstuhl -- Leibniz-Zentrum f{\"u}r Informatik},
  year      = {2026},
  doi       = {10.4230/LIPIcs.CPM.2026.2}
}

@inproceedings{Kopelowitz12,
  author    = {Tsvi Kopelowitz},
  title     = {On-Line Indexing for General Alphabets via Predecessor Queries
               on Subsets of an Ordered List},
  booktitle = {Proc. 53rd Annual IEEE Symposium on Foundations of Computer
               Science (FOCS)},
  pages     = {283--292},
  publisher = {IEEE},
  year      = {2012},
  doi       = {10.1109/FOCS.2012.79}
}

@inproceedings{Weiner73,
  author    = {Peter Weiner},
  title     = {Linear Pattern Matching Algorithms},
  booktitle = {14th Annual Symposium on Switching and Automata Theory (SWAT)},
  pages     = {1--11},
  publisher = {IEEE},
  year      = {1973}
}

@article{McCreight76,
  author  = {Edward M. McCreight},
  title   = {A Space-Economical Suffix Tree Construction Algorithm},
  journal = {Journal of the ACM},
  volume  = {23},
  number  = {2},
  pages   = {262--272},
  year    = {1976}
}

@article{FarachFM00,
  author  = {Martin Farach-Colton and Paolo Ferragina and S. Muthukrishnan},
  title   = {On the Sorting-Complexity of Suffix Tree Construction},
  journal = {Journal of the ACM},
  volume  = {47},
  number  = {6},
  pages   = {987--1011},
  year    = {2000}
}

@article{ManberM93,
  author  = {Udi Manber and Gene Myers},
  title   = {Suffix Arrays: A New Method for On-Line String Searches},
  journal = {SIAM Journal on Computing},
  volume  = {22},
  number  = {5},
  pages   = {935--948},
  year    = {1993}
}

@article{KarkkainenSB06,
  author  = {Juha K{\"a}rkk{\"a}inen and Peter Sanders and Stefan Burkhardt},
  title   = {Linear Work Suffix Array Construction},
  journal = {Journal of the ACM},
  volume  = {53},
  number  = {6},
  pages   = {918--936},
  year    = {2006}
}

@article{KimSPP05,
  author  = {Dong Kyue Kim and Jeong Seop Sim and Heejin Park and Kunsoo Park},
  title   = {Constructing Suffix Arrays in Linear Time},
  journal = {Journal of Discrete Algorithms},
  volume  = {3},
  number  = {2--4},
  pages   = {126--142},
  year    = {2005}
}

@article{KoA05,
  author  = {Pang Ko and Srinivas Aluru},
  title   = {Space Efficient Linear Time Construction of Suffix Arrays},
  journal = {Journal of Discrete Algorithms},
  volume  = {3},
  number  = {2--4},
  pages   = {143--156},
  year    = {2005}
}

@article{GrossiV05,
  author  = {Roberto Grossi and Jeffrey Scott Vitter},
  title   = {Compressed Suffix Arrays and Suffix Trees with Applications to Text Indexing and String Matching},
  journal = {SIAM Journal on Computing},
  volume  = {35},
  number  = {2},
  pages   = {378--407},
  year    = {2005}
}

@article{FerraginaM05,
  author  = {Paolo Ferragina and Giovanni Manzini},
  title   = {Indexing Compressed Text},
  journal = {Journal of the ACM},
  volume  = {52},
  number  = {4},
  pages   = {552--581},
  year    = {2005}
}

@article{NavarroM07,
  author  = {Gonzalo Navarro and Veli M{\"a}kinen},
  title   = {Compressed Full-Text Indexes},
  journal = {ACM Computing Surveys},
  volume  = {39},
  number  = {1},
  pages   = {2},
  year    = {2007}
}

@inproceedings{AlstrupHR98,
  author    = {Stephen Alstrup and Thore Husfeldt and Theis Rauhe},
  title     = {Marked Ancestor Problems},
  booktitle = {Proc. 39th Annual Symposium on Foundations of Computer Science (FOCS)},
  pages     = {534--543},
  publisher = {IEEE},
  year      = {1998},
  doi       = {10.1109/SFCS.1998.743504}
}

@inproceedings{AlstrupBR00,
  author    = {Stephen Alstrup and Gerth St{\o}lting Brodal and Theis Rauhe},
  title     = {Pattern Matching in Dynamic Texts},
  booktitle = {Proc. 11th Annual ACM-SIAM Symposium on Discrete Algorithms (SODA)},
  pages     = {819--828},
  publisher = {SIAM},
  year      = {2000}
}

@inproceedings{ColeGL04,
  author    = {Richard Cole and Lee-Ad Gottlieb and Moshe Lewenstein},
  title     = {Dictionary Matching and Indexing with Errors and Don't Cares},
  booktitle = {Proc. 36th Annual ACM Symposium on Theory of Computing (STOC)},
  pages     = {91--100},
  publisher = {ACM},
  year      = {2004},
  doi       = {10.1145/1007352.1007374}
}

@article{ColeH05,
  author    = {Richard Cole and Ramesh Hariharan},
  title     = {Dynamic {LCA} Queries on Trees},
  journal   = {SIAM Journal on Computing},
  volume    = {34},
  number    = {4},
  pages     = {894--923},
  year      = {2005}
}

@inproceedings{DemaineMPP06,
  author    = {Erik D. Demaine and Friedhelm {Meyer auf der Heide} and Rasmus Pagh and Mihai P{\u{a}}tra{\c{s}}cu},
  title     = {De Dictionariis Dynamicis Pauco Spatio Utentibus (lat. On Dynamic Dictionaries Using Little Space)},
  booktitle = {Proc. 7th Latin American Symposium on Theoretical Informatics (LATIN)},
  series    = {Lecture Notes in Computer Science},
  pages     = {349--361},
  publisher = {Springer},
  year      = {2006},
  doi       = {10.1007/11682462_34}
}

@inproceedings{DietzS87,
  author    = {Paul F. Dietz and Daniel D. Sleator},
  title     = {Two Algorithms for Maintaining Order in a List},
  booktitle = {Proc. 19th Annual ACM Symposium on Theory of Computing (STOC)},
  pages     = {365--372},
  publisher = {ACM},
  year      = {1987},
  doi       = {10.1145/28395.28434}
}

@article{Ferragina97,
  author    = {Paolo Ferragina},
  title     = {Dynamic Text Indexing under String Updates},
  journal   = {Journal of Algorithms},
  volume    = {22},
  number    = {2},
  pages     = {296--328},
  year      = {1997}
}

@article{FerraginaG99,
  author    = {Paolo Ferragina and Roberto Grossi},
  title     = {The String {B}-tree: A New Data Structure for String Search in External Memory and Its Applications},
  journal   = {Journal of the ACM},
  volume    = {46},
  number    = {2},
  pages     = {236--280},
  year      = {1999}
}

@article{GagieNP20,
  author    = {Travis Gagie and Gonzalo Navarro and Nicola Prezza},
  title     = {Fully Functional Suffix Trees and Optimal Text Searching in {BWT}-Runs Bounded Space},
  journal   = {Journal of the ACM},
  volume    = {67},
  number    = {1},
  year      = {2020},
  doi       = {10.1145/3375890}
}

@inproceedings{GuFB94,
  author    = {Ming Gu and Martin Farach and Richard Beigel},
  title     = {An Efficient Algorithm for Dynamic Text Indexing},
  booktitle = {Proc. 5th Annual ACM-SIAM Symposium on Discrete Algorithms (SODA)},
  pages     = {697--704},
  publisher = {SIAM},
  year      = {1994}
}

@inproceedings{KempaK22,
  author    = {Dominik Kempa and Tomasz Kociumaka},
  title     = {Dynamic Suffix Array with Polylogarithmic Queries and Updates},
  booktitle = {Proc. 54th Annual ACM SIGACT Symposium on Theory of Computing (STOC)},
  pages     = {1657--1670},
  publisher = {ACM},
  year      = {2022},
  doi       = {10.1145/3519935.3520061}
}

@inproceedings{Muthukrishnan02,
  author    = {S. Muthukrishnan},
  title     = {Efficient Algorithms for Document Retrieval Problems},
  booktitle = {Proc. 13th Annual ACM-SIAM Symposium on Discrete Algorithms (SODA)},
  pages     = {657--666},
  publisher = {SIAM},
  year      = {2002}
}

@inproceedings{PatrascuT06,
  author    = {Mihai P{\u{a}}tra{\c{s}}cu and Mikkel Thorup},
  title     = {Time-Space Trade-Offs for Predecessor Search},
  booktitle = {Proc. 38th Annual ACM Symposium on Theory of Computing (STOC)},
  pages     = {232--240},
  publisher = {ACM},
  year      = {2006},
  doi       = {10.1145/1132516.1132551}
}

@inproceedings{SahinalpV96,
  author    = {S{\"u}leyman Cenk Sahinalp and Uzi Vishkin},
  title     = {Efficient Approximate and Dynamic Matching of Patterns Using a Labeling Paradigm},
  booktitle = {Proc. 37th Annual Symposium on Foundations of Computer Science (FOCS)},
  pages     = {320--328},
  publisher = {IEEE},
  year      = {1996}
}

@article{Willard92,
  author    = {Dan E. Willard},
  title     = {A Density Control Algorithm for Doing Insertions and Deletions in a Sequentially Ordered File in a Good Worst-Case Time},
  journal   = {Information and Computation},
  volume    = {97},
  number    = {2},
  pages     = {150--204},
  year      = {1992}
}

\end{document}